\documentclass[a4paper,10pt]{article} 

\usepackage{units}
\usepackage{amsmath}
\usepackage{amssymb}
\usepackage[numbers,sort&compress]{natbib}
\usepackage{hyperref}
\usepackage{url}
\usepackage{multirow}
\usepackage{enumitem}
\usepackage{xcolor}
\usepackage{caption}
\usepackage{geometry}
\usepackage{graphicx, color}

\newcommand{\xmaxM}{X_{\mathrm{max}}}
\newcommand{\fmuM}{F_{\mu}}

\begin{document}

\begin{center}
\textbf{\LARGE{Searches for Ultra-High-Energy Photons\\at the Pierre Auger Observatory}}\\[10mm]
\large{The Pierre Auger Collaboration}
\end{center}
\normalsize{
P.~Abreu$^{71}$,
M.~Aglietta$^{53,51}$,
I.~Allekotte$^{1}$,
K.~Almeida Cheminant$^{69}$,
A.~Almela$^{8,12}$,
J.~Alvarez-Mu\~niz$^{78}$,
J.~Ammerman Yebra$^{78}$,
G.A.~Anastasi$^{53,51}$,
L.~Anchordoqui$^{85}$,
B.~Andrada$^{8}$
% ...
}
\small{
\begin{description}[labelsep=0.2em,align=right,labelwidth=0.7em,labelindent=0em,leftmargin=2em,noitemsep]
\item[$^{1}$] Centro At\'omico Bariloche and Instituto Balseiro (CNEA-UNCuyo-CONICET), San Carlos de Bariloche, Argentina
\item[$^{2}$] Centro de Investigaciones en L\'aseres y Aplicaciones, CITEDEF and CONICET, Villa Martelli, Argentina
\item[$^{3}$] Departamento de F\'\i{}sica and Departamento de Ciencias de la Atm\'osfera y los Oc\'eanos, FCEyN, Universidad de Buenos Aires and CONICET, Buenos Aires, Argentina
\item[$^{4}$] IFLP, Universidad Nacional de La Plata and CONICET, La Plata, Argentina
\item[$^{5}$] Instituto de Astronom\'\i{}a y F\'\i{}sica del Espacio (IAFE, CONICET-UBA), Buenos Aires, Argentina
\item[$^{6}$] Instituto de F\'\i{}sica de Rosario (IFIR) -- CONICET/U.N.R.\ and Facultad de Ciencias Bioqu\'\i{}micas y Farmac\'euticas U.N.R., Rosario, Argentina
\item[$^{7}$] Instituto de Tecnolog\'\i{}as en Detecci\'on y Astropart\'\i{}culas (CNEA, CONICET, UNSAM), and Universidad Tecnol\'ogica Nacional -- Facultad Regional Mendoza (CONICET/CNEA), Mendoza, Argentina
\item[$^{8}$] Instituto de Tecnolog\'\i{}as en Detecci\'on y Astropart\'\i{}culas (CNEA, CONICET, UNSAM), Buenos Aires, Argentina
\item[$^{9}$] International Center of Advanced Studies and Instituto de Ciencias F\'\i{}sicas, ECyT-UNSAM and CONICET, Campus Miguelete -- San Mart\'\i{}n, Buenos Aires, Argentina
\item[$^{10}$] Observatorio Pierre Auger, Malarg\"ue, Argentina
\item[$^{11}$] Observatorio Pierre Auger and Comisi\'on Nacional de Energ\'\i{}a At\'omica, Malarg\"ue, Argentina
\item[$^{12}$] Universidad Tecnol\'ogica Nacional -- Facultad Regional Buenos Aires, Buenos Aires, Argentina
\item[$^{13}$] University of Adelaide, Adelaide, S.A., Australia
\item[$^{14}$] Universit\'e Libre de Bruxelles (ULB), Brussels, Belgium
\item[$^{15}$] Vrije Universiteit Brussels, Brussels, Belgium
\item[$^{16}$] Centro Federal de Educa\c{c}\~ao Tecnol\'ogica Celso Suckow da Fonseca, Nova Friburgo, Brazil
\item[$^{17}$] Instituto Federal de Educa\c{c}\~ao, Ci\^encia e Tecnologia do Rio de Janeiro (IFRJ), Brazil
\item[$^{18}$] Universidade de S\~ao Paulo, Escola de Engenharia de Lorena, Lorena, SP, Brazil
\item[$^{19}$] Universidade de S\~ao Paulo, Instituto de F\'\i{}sica de S\~ao Carlos, S\~ao Carlos, SP, Brazil
\item[$^{20}$] Universidade de S\~ao Paulo, Instituto de F\'\i{}sica, S\~ao Paulo, SP, Brazil
\item[$^{21}$] Universidade Estadual de Campinas, IFGW, Campinas, SP, Brazil
\item[$^{22}$] Universidade Estadual de Feira de Santana, Feira de Santana, Brazil
\item[$^{23}$] Universidade Federal do ABC, Santo Andr\'e, SP, Brazil
\item[$^{24}$] Universidade Federal do Paran\'a, Setor Palotina, Palotina, Brazil
\item[$^{25}$] Universidade Federal do Rio de Janeiro, Instituto de F\'\i{}sica, Rio de Janeiro, RJ, Brazil
\item[$^{26}$] Universidade Federal do Rio de Janeiro (UFRJ), Observat\'orio do Valongo, Rio de Janeiro, RJ, Brazil
\item[$^{27}$] Universidade Federal Fluminense, EEIMVR, Volta Redonda, RJ, Brazil
\item[$^{28}$] Universidad de Medell\'\i{}n, Medell\'\i{}n, Colombia
\item[$^{29}$] Universidad Industrial de Santander, Bucaramanga, Colombia
\item[$^{30}$] Charles University, Faculty of Mathematics and Physics, Institute of Particle and Nuclear Physics, Prague, Czech Republic
\item[$^{31}$] Institute of Physics of the Czech Academy of Sciences, Prague, Czech Republic
\item[$^{32}$] Palacky University, RCPTM, Olomouc, Czech Republic
\item[$^{33}$] CNRS/IN2P3, IJCLab, Universit\'e Paris-Saclay, Orsay, France
\item[$^{34}$] Laboratoire de Physique Nucl\'eaire et de Hautes Energies (LPNHE), Sorbonne Universit\'e, Universit\'e de Paris, CNRS-IN2P3, Paris, France
\item[$^{35}$] Univ.\ Grenoble Alpes, CNRS, Grenoble Institute of Engineering Univ.\ Grenoble Alpes, LPSC-IN2P3, 38000 Grenoble, France
\item[$^{36}$] Universit\'e Paris-Saclay, CNRS/IN2P3, IJCLab, Orsay, France
\item[$^{37}$] Bergische Universit\"at Wuppertal, Department of Physics, Wuppertal, Germany
\item[$^{38}$] Karlsruhe Institute of Technology (KIT), Institute for Experimental Particle Physics, Karlsruhe, Germany
\item[$^{39}$] Karlsruhe Institute of Technology (KIT), Institut f\"ur Prozessdatenverarbeitung und Elektronik, Karlsruhe, Germany
\item[$^{40}$] Karlsruhe Institute of Technology (KIT), Institute for Astroparticle Physics, Karlsruhe, Germany
\item[$^{41}$] RWTH Aachen University, III.\ Physikalisches Institut A, Aachen, Germany
\item[$^{42}$] Universit\"at Hamburg, II.\ Institut f\"ur Theoretische Physik, Hamburg, Germany
\item[$^{43}$] Universit\"at Siegen, Department Physik -- Experimentelle Teilchenphysik, Siegen, Germany
\item[$^{44}$] Gran Sasso Science Institute, L'Aquila, Italy
\item[$^{45}$] INFN Laboratori Nazionali del Gran Sasso, Assergi (L'Aquila), Italy
\item[$^{46}$] INFN, Sezione di Catania, Catania, Italy
\item[$^{47}$] INFN, Sezione di Lecce, Lecce, Italy
\item[$^{48}$] INFN, Sezione di Milano, Milano, Italy
\item[$^{49}$] INFN, Sezione di Napoli, Napoli, Italy
\item[$^{50}$] INFN, Sezione di Roma ``Tor Vergata'', Roma, Italy
\item[$^{51}$] INFN, Sezione di Torino, Torino, Italy
\item[$^{52}$] Istituto di Astrofisica Spaziale e Fisica Cosmica di Palermo (INAF), Palermo, Italy
\item[$^{53}$] Osservatorio Astrofisico di Torino (INAF), Torino, Italy
\item[$^{54}$] Politecnico di Milano, Dipartimento di Scienze e Tecnologie Aerospaziali , Milano, Italy
\item[$^{55}$] Universit\`a del Salento, Dipartimento di Matematica e Fisica ``E.\ De Giorgi'', Lecce, Italy
\item[$^{56}$] Universit\`a dell'Aquila, Dipartimento di Scienze Fisiche e Chimiche, L'Aquila, Italy
\item[$^{57}$] Universit\`a di Catania, Dipartimento di Fisica e Astronomia ``Ettore Majorana``, Catania, Italy
\item[$^{58}$] Universit\`a di Milano, Dipartimento di Fisica, Milano, Italy
\item[$^{59}$] Universit\`a di Napoli ``Federico II'', Dipartimento di Fisica ``Ettore Pancini'', Napoli, Italy
\item[$^{60}$] Universit\`a di Palermo, Dipartimento di Fisica e Chimica ''E.\ Segr\`e'', Palermo, Italy
\item[$^{61}$] Universit\`a di Roma ``Tor Vergata'', Dipartimento di Fisica, Roma, Italy
\item[$^{62}$] Universit\`a Torino, Dipartimento di Fisica, Torino, Italy
\item[$^{63}$] Benem\'erita Universidad Aut\'onoma de Puebla, Puebla, M\'exico
\item[$^{64}$] Unidad Profesional Interdisciplinaria en Ingenier\'\i{}a y Tecnolog\'\i{}as Avanzadas del Instituto Polit\'ecnico Nacional (UPIITA-IPN), M\'exico, D.F., M\'exico
\item[$^{65}$] Universidad Aut\'onoma de Chiapas, Tuxtla Guti\'errez, Chiapas, M\'exico
\item[$^{66}$] Universidad Michoacana de San Nicol\'as de Hidalgo, Morelia, Michoac\'an, M\'exico
\item[$^{67}$] Universidad Nacional Aut\'onoma de M\'exico, M\'exico, D.F., M\'exico
\item[$^{68}$] Universidad Nacional de San Agustin de Arequipa, Facultad de Ciencias Naturales y Formales, Arequipa, Peru
\item[$^{69}$] Institute of Nuclear Physics PAN, Krakow, Poland
\item[$^{70}$] University of \L{}\'od\'z, Faculty of High-Energy Astrophysics,\L{}\'od\'z, Poland
\item[$^{71}$] Laborat\'orio de Instrumenta\c{c}\~ao e F\'\i{}sica Experimental de Part\'\i{}culas -- LIP and Instituto Superior T\'ecnico -- IST, Universidade de Lisboa -- UL, Lisboa, Portugal
\item[$^{72}$] ``Horia Hulubei'' National Institute for Physics and Nuclear Engineering, Bucharest-Magurele, Romania
\item[$^{73}$] Institute of Space Science, Bucharest-Magurele, Romania
\item[$^{74}$] University Politehnica of Bucharest, Bucharest, Romania
\item[$^{75}$] Center for Astrophysics and Cosmology (CAC), University of Nova Gorica, Nova Gorica, Slovenia
\item[$^{76}$] Experimental Particle Physics Department, J.\ Stefan Institute, Ljubljana, Slovenia
\item[$^{77}$] Universidad de Granada and C.A.F.P.E., Granada, Spain
\item[$^{78}$] Instituto Galego de F\'\i{}sica de Altas Enerx\'\i{}as (IGFAE), Universidade de Santiago de Compostela, Santiago de Compostela, Spain
\item[$^{79}$] IMAPP, Radboud University Nijmegen, Nijmegen, The Netherlands
\item[$^{80}$] Nationaal Instituut voor Kernfysica en Hoge Energie Fysica (NIKHEF), Science Park, Amsterdam, The Netherlands
\item[$^{81}$] Stichting Astronomisch Onderzoek in Nederland (ASTRON), Dwingeloo, The Netherlands
\item[$^{82}$] Universiteit van Amsterdam, Faculty of Science, Amsterdam, The Netherlands
\item[$^{83}$] Case Western Reserve University, Cleveland, OH, USA
\item[$^{84}$] Colorado School of Mines, Golden, CO, USA
\item[$^{85}$] Department of Physics and Astronomy, Lehman College, City University of New York, Bronx, NY, USA
\item[$^{86}$] Michigan Technological University, Houghton, MI, USA
\item[$^{87}$] New York University, New York, NY, USA
\item[$^{88}$] University of Chicago, Enrico Fermi Institute, Chicago, IL, USA
\item[$^{89}$] University of Delaware, Department of Physics and Astronomy, Bartol Research Institute, Newark, DE, USA
\item[$^{90}$] University of Wisconsin-Madison, Department of Physics and WIPAC, Madison, WI, USA
\item[] -----
\item[$^{a}$] Louisiana State University, Baton Rouge, LA, USA
\item[$^{b}$] also at University of Bucharest, Physics Department, Bucharest, Romania
\item[$^{c}$] School of Physics and Astronomy, University of Leeds, Leeds, United Kingdom
\item[$^{d}$] Fermi National Accelerator Laboratory, Fermilab, Batavia, IL, USA
\item[$^{e}$] now at Graduate School of Science, Osaka Metropolitan University, Osaka, Japan
\item[$^{f}$] Max-Planck-Institut f\"ur Radioastronomie, Bonn, Germany
\item[$^{g}$] Colorado State University, Fort Collins, CO, USA
\item[$^{h}$] Pennsylvania State University, University Park, PA, USA\vspace*{3mm}
\item Correspondence: \url{spokespersons@auger.org}
\end{description}
}\vspace*{2mm}

\abstract{The Pierre Auger Observatory, being the largest air-shower experiment in the world, offers an unprecedented exposure to neutral particles at the highest energies. Since the start of data taking more than 18 years ago, various searches for ultra-high-energy (UHE, \texorpdfstring{$E\gtrsim\unit[10^{17}]{eV}$}{E>1017ev}) photons have been performed: either for a diffuse flux of UHE photons, for point sources of UHE photons or for UHE photons associated with transient events like gravitational wave events. In the present paper, we summarize these searches and review the current results obtained using the wealth of data collected by the Pierre Auger Observatory.}

\vspace*{1mm}


\section{Introduction}
\label{sec:intro}

The search for neutral particles---in particular photons and neutrinos---of cosmic origin at the highest energies has been for many years one of the major scientific objectives of the Pierre Auger Observatory. From the theory side, such searches are well-motivated: many models for the origin of ultra-high-energy (UHE) cosmic rays predict at least some neutral particles as by-products, either directly at the sources or during the propagation through the Universe (see, e.g., \cite{Gelmini:2005wu,Kampert:2011hkm,Bobrikova:2021kuj,Berat2021,Gelmini:2022sti}). In fact, even though no UHE photons have been unambiguously identified so far, the upper limits on their incoming flux have been already used to severely constrain so-called top-down models for the origin of UHE cosmic rays involving e.g. topological defects or super-heavy dark matter (see, e.g., \cite{Aloisio:2015lva,Anchordoqui:2021crl,PierreAuger:2022hfz,PierreAuger:2022swm}). In addition, the recent observations of photons with energies up to $\unit[10^{15}]{eV}$~\cite{LhaasoNature} further motivate searches for photons at even higher energies. An observation of such photons would also be key in completing the multi-messenger approach aimed at understanding the most extreme processes in the Universe, taking advantage of the fact that neutral particles directly point back at their production site. However, one has to take into account that UHE photons, unlike neutrinos, interact with the background photon fields permeating the Universe, reducing their attenuation length to about $\unit[30]{kpc}$ around $\unit[10^{15}]{eV}$, which increases to the order of $\unit[10]{Mpc}$ around $\unit[10^{19}]{eV}$~\cite{Risse:2007sd}. In the present paper, we review the current state of such searches at the Pierre Auger Observatory. After a short introduction addressing the specificities of air showers initiated by photons (Sec.~\ref{sec:photonshowers}), we briefly describe the Pierre Auger Observatory (Sec.~\ref{sec:auger}). We then focus first on the searches for a diffuse flux of UHE photons using the different detector systems of the Observatory~(Sec.~\ref{sec:searchesdiffuse}), before we describe the searches for UHE photons from point sources and transient events (Sec.~\ref{sec:searchespointsources}). We close with a short outlook (Sec.~\ref{sec:outlook}), discussing the ongoing detector upgrade of the Pierre Auger Observatory, dubbed AugerPrime.


\section{Photon-induced air showers}
\label{sec:photonshowers}

\begin{figure}[t]
   \centering
   \includegraphics[width=0.6\textwidth]{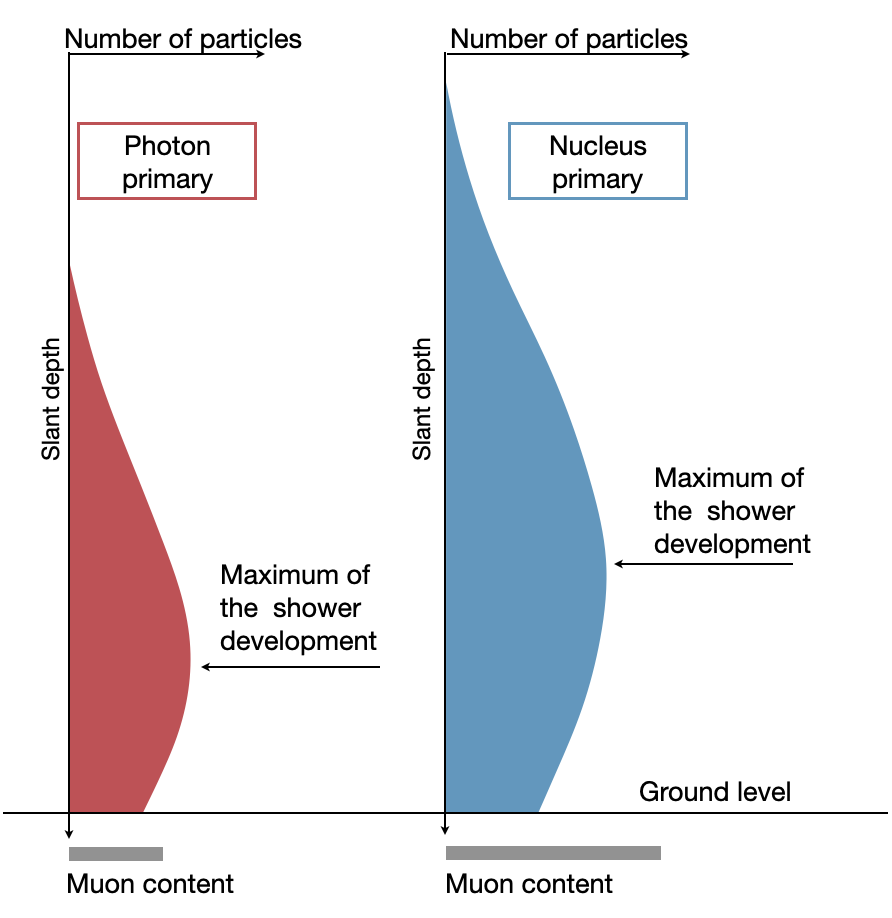}
	\caption{Schematic depiction of the main differences between photon-induced air showers and those initiated by primary nuclei (protons or heavier nuclei).}
	\label{fig:photonshowers}
\end{figure}

When a UHE photon enters the Earth's atmosphere, it may interact with a particle from the atmosphere, for example a nitrogen nucleus, and induce an extensive air shower, much in the same way as a charged cosmic ray does. Hence, a cosmic-ray observatory detecting air showers is also, by construction, a photon observatory---and even a neutrino observatory, highlighting the importance of such observatories for multi-messenger astrophysics. In fact, since the incoming flux of  UHE cosmic particles is so low (on the order of one particle per square kilometer and year and less), measuring the extensive air showers they initiate in the atmosphere with large detector arrays on ground is the only way to efficiently detect them. The challenge lies in distinguishing air showers induced by photons from the vast background of air showers that are initiated by charged cosmic rays, i.e., protons and heavier nuclei (for a review, see, e.g.,~\cite{Risse:2007sd}). The two main differences between photon- and nucleus-induced air showers are shown schematically in Fig.~\ref{fig:photonshowers}. The longitudinal development of an air shower, as a function of the slant depth $X$, is delayed for a primary photon with respect to primary nuclei, due to the lower multiplicity of electromagnetic interactions (compared to hadronic interactions) that dominate in a photon-induced air shower. The maximum of the shower development within the atmosphere, $\xmaxM$, is reached later. For example, at a primary energy of $\unit[10^{19}]{eV}$, the difference is about $\unit[200]{g\,cm^{-2}}$. Since the mean free path for photo-nuclear interactions is much larger than the radiation length, only a small fraction of the electromagnetic component in a photon-induced shower is transferred to the hadronic component and subsequently to the muonic component. Showers induced by photons are thus characterized by a lower number of muons. On average, simulations show that photon-induced showers have nearly one order of magnitude less muons than those initiated by protons or nuclei of the same primary energy. In one way or another, all searches for UHE photons using air-shower data exploit these two key differences: $\xmaxM$, for example, can be directly measured using the air-fluorescence technique. The number of muons cannot yet be directly measured using the current detector systems of the Pierre Auger Observatory. However, one can measure the lateral distribution of secondary particles from the air shower at ground level, which depends on both the number of muons and the longitudinal development. In particular, the steepness of the lateral distribution is sensitive to the type of the primary particle.\\

The difference in $\xmaxM$ between photon- and proton-induced air showers
is amplified by the Landau-Pomeranchuk-Migdal (LPM) effect~\cite{Landau:1953um,Migdal:1956tc}, i.e., the suppression of the bremsstrahlung and pair production cross sections at high energies. In addition, the preshower effect~\cite{Erber:1966vv,Mcbreen:1981yc,Homola:2006wf} also has to be taken into account: depending on its incident direction with respect to the local geomagnetic field, a UHE photon can initiate an electromagnetic cascade in the Earth's magnetic field even before entering the atmosphere---a preshower. The observed air shower is then a superposition of the individual cascades of the photons and electrons/positrons from the preshower, leading on average to a smaller measured $\xmaxM$ than for non-preshowering photons of the same primary energy. 


\section{The Pierre Auger Observatory}
\label{sec:auger}

\begin{figure}[t]
    \centering
    \includegraphics[width=\textwidth]{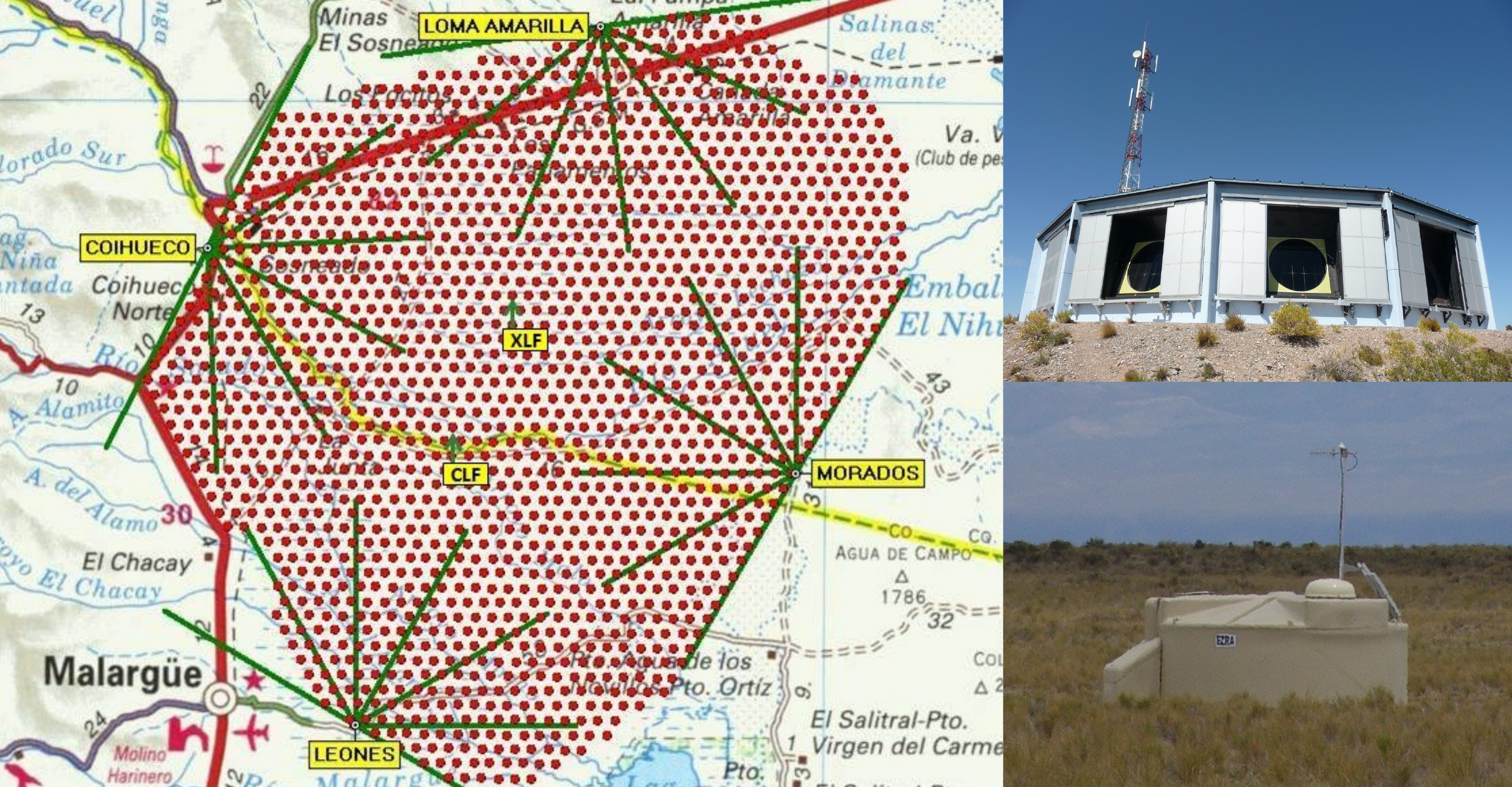}
    \caption{Left: map of the Pierre Auger Observatory~\cite{ThePierreAuger:2015rma}; each dot represents one SD station; also shown are the four FD sites at the border of the SD array. Top right: the fluorescence telescopes at the FD site Los Leones; even though the picture was taken during the daytime, the shutters were had been opened for maintenance. Bottom right: a single SD station in the \textit{Pampa Amarilla}.}
    \label{fig:auger}
\end{figure}

The Pierre Auger Observatory~\citep{ThePierreAuger:2015rma}, located near the town of Malarg\"ue in the Argentinian \textit{Pampa Amarilla}, is the largest cosmic-ray observatory to date, offering an unprecedented exposure to UHE photons. A key feature of the Pierre Auger Observatory is the hybrid concept, combining a Surface Detector array (SD) with a Fluorescence Detector (FD), see Fig.~\ref{fig:auger}. The SD consists of 1600 water-Cherenkov detectors arranged on a triangular grid with a spacing of $\unit[1500]{m}$, covering a total area of about $\unit[3000]{km^2}$. The SD is overlooked by 24 fluorescence telescopes, located at four sites at the border of the array. The SD samples the lateral shower profile at ground level, i.e., the distribution of particles as a function of the distance from the shower axis, with a duty cycle of $\unit[{\sim}100]{\%}$, while the FD records the longitudinal shower development in the atmosphere above the SD\@. The FD can only be operated in clear, moonless nights, reducing the duty cycle to $\unit[{\sim}15]{\%}$. Through combining measurements from both detector systems in hybrid events, a superior accuracy of the air-shower reconstruction can be achieved than with just one system~\cite{PierreAuger:2003wmb}. In the western part of the SD array, 50 additional SD stations have been placed between the existing SD stations, forming a sub-array with a spacing of $\unit[750]{m}$ and covering a total area of about $\unit[27.5]{km^2}$. With this sub-array, air showers of lower primary energy (below $\unit[10^{18}]{eV}$) with a smaller footprint on ground can be measured. To allow also for hybrid measurements in this energy range, where air showers develop above the field of view of the standard FD telescopes, three additional High-Elevation Auger Telescopes (HEAT) have been installed at the FD site Coihueco, overlooking the $\unit[750]{m}$ SD array. The HEAT telescopes operate in the range of elevation angles from $30^\circ$ to $60^\circ$, complementing the Coihueco telescopes operating in the $0^\circ$ to $30^\circ$ range. The combination of the data from both HEAT and Coihueco (``HeCo'' data) enables fluorescence measurements of air showers over a large range of elevation angles.

\section{Searches for a diffuse flux of UHE photons}
\label{sec:searchesdiffuse}

First, we focus on the searches for a diffuse---i.e., direction-independent, unresolved---flux of photons. At the Pierre Auger Observatory, such searches have been performed in the past using hybrid data (with the analysis based on $\xmaxM$ only~\cite{PierreAuger:2006lld,PierreAuger:2009yle}, or based on a combination of $\xmaxM$ and additional SD-related quantities~\cite{Aab:2016agp}) as well as SD-only data~\cite{PierreAuger:2007hjd}. In the following, we briefly summarize the three most up-to-date publications~\cite{PierreAuger:2022uwd,PierreAuger:2021mjh,PierreAuger:2022abc} in different energy ranges and the corresponding results.


\subsection{A search for photons with energies above \texorpdfstring{$\unit[2{\times}10^{17}]{eV}$}{2x1017 eV} using hybrid data from the low-energy extensions of the Pierre Auger Observatory}
\label{sec:heco}

We first discuss the photon search targeting the lowest energy range, which starts from $\unit[2{\times}10^{17}]{eV}$~\cite{PierreAuger:2022uwd}. In this energy range, data from the low-energy extensions of the Pierre Auger Observatory, i.e., from the $\unit[750]{m}$ SD array combined with HeCo data, can be used to efficiently search for photons. Three observables are used in the analysis: $\xmaxM$, measured directly with the fluorescence telescopes, is used together with the SD quantities $S_b$ and $N_\text{stations}$. $S_b$~\cite{Ros:2011zg} is defined through 
\begin{equation}
   S_b = \sum_i S_i \times \left(\frac{R_i}{\unit[1000]{m}}\right)^{b},
\end{equation}
where $S_i$ denotes the measured signal in the $i$-th SD station at a perpendicular distance $R_i$ to the shower axis and the parameter $b$ has been chosen as $b=4$ to optimize the photon-hadron separation. By construction, $S_b$ is sensitive to the lateral distribution, which in turn depends on the depth of the air-shower development in the atmosphere and the number of muons, as stated in Sec.~\ref{sec:photonshowers}. The third observable, $N_\text{stations}$ is the number of triggered SD stations, as it has been shown previously that it can significantly improve the overall performance of the analysis~\cite{Aab:2016agp}.\\

These three quantities are combined in a multivariate analysis (MVA) using the Boosted Decision Tree (BDT) method. To take into account energy and zenith angle dependencies, the photon energy $E_\gamma$ (defined as the calorimetric energy obtained through the integration of the longitudinal profile plus a missing-energy correction of $\unit[1]{\%}$ appropriate for primary photons~\cite{Aab:2016agp}) and the reconstructed zenith angle $\theta$ are also included in the MVA. A large sample of simulated events has been used to study the photon/hadron separation of the observables mentioned before, to train the multivariate analysis, and to evaluate its performance. For these samples, both primary photons (as a signal sample) and primary protons (as a conservative, ``worst-case'' assumption for the background sample) have been simulated using CORSIKA and the Auger Offline Software Framework. The analysis is eventually  applied to hybrid data collected by the Coihueco and HEAT telescopes and the $\unit[750]{m}$ SD array between 1 June 2010 and 31 December 2015, in total more than $500{,}000$ events. A number of selection criteria is applied to both the data sample and the simulated samples to select only well-reconstructed, reliable events. These selection criteria are described in detail in~\cite{PierreAuger:2022uwd}. After all criteria have been applied, $2{,}204$ events remain in the data sample with a photon energy $E_\gamma$ above $\unit[2{\times}10^{17}]{eV}$.\\

\begin{figure}[t]
   \centering
   \includegraphics[width=0.49\textwidth]{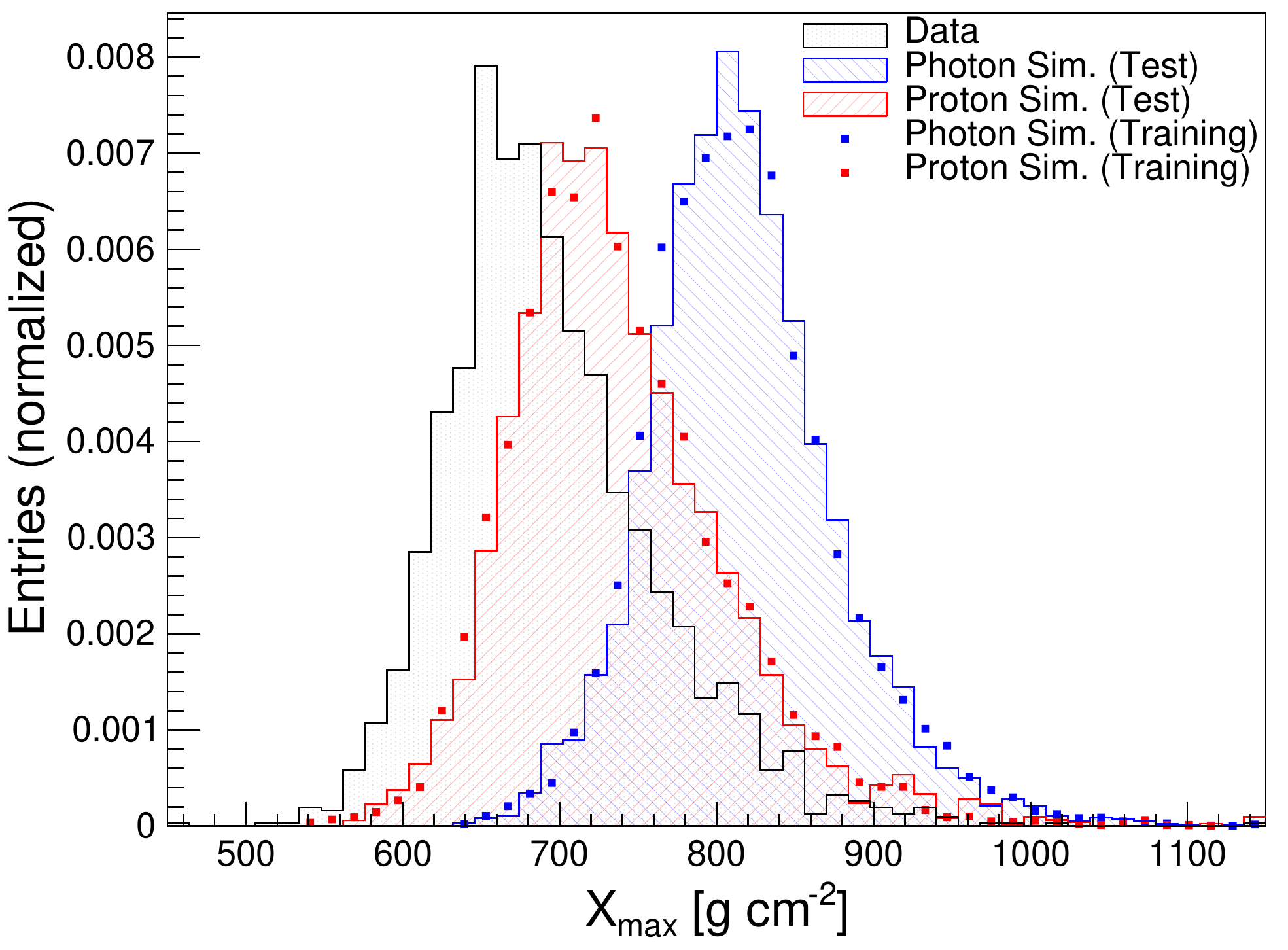}
   \includegraphics[width=0.49\textwidth]{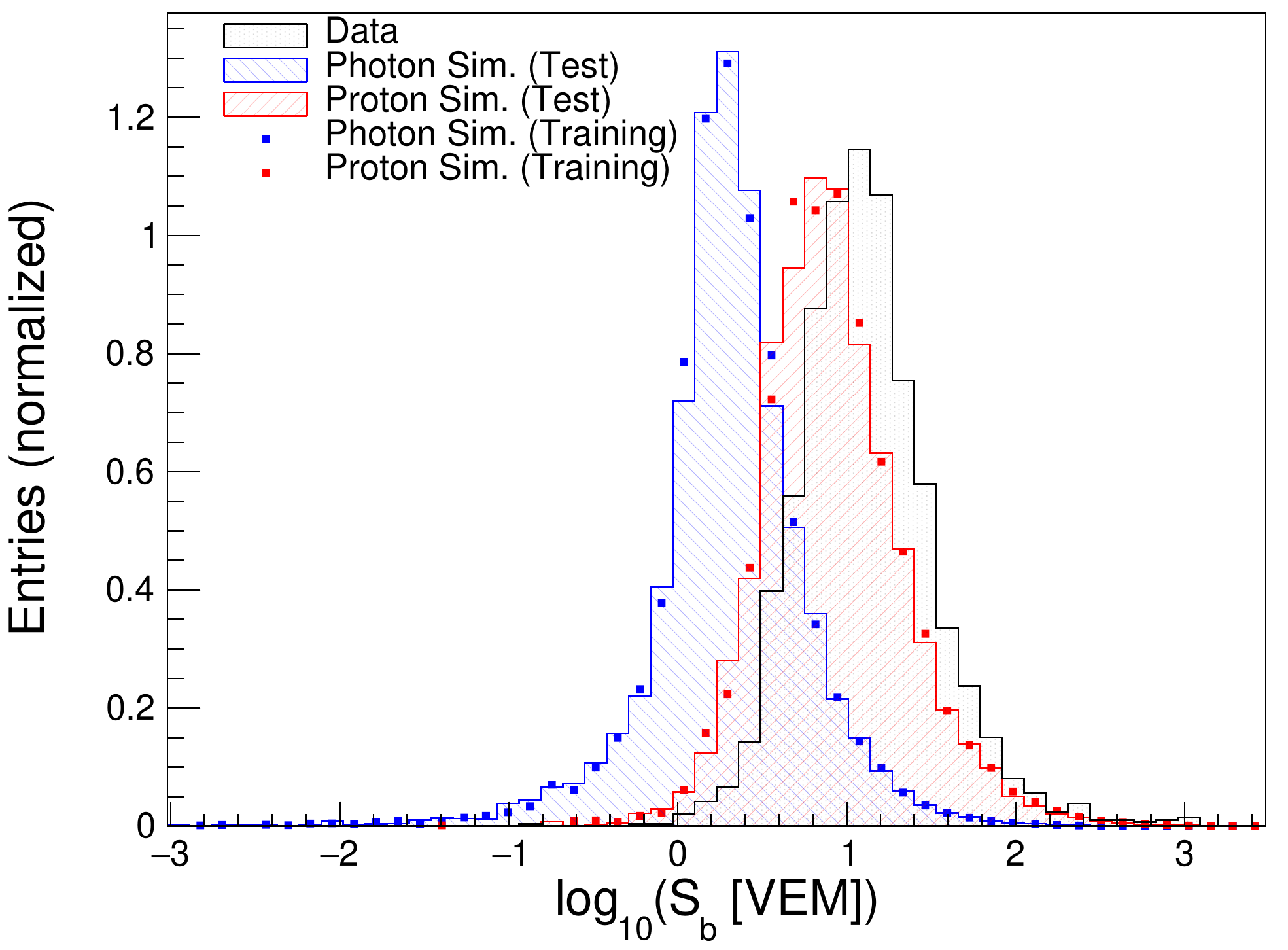}
   \includegraphics[width=0.49\textwidth]{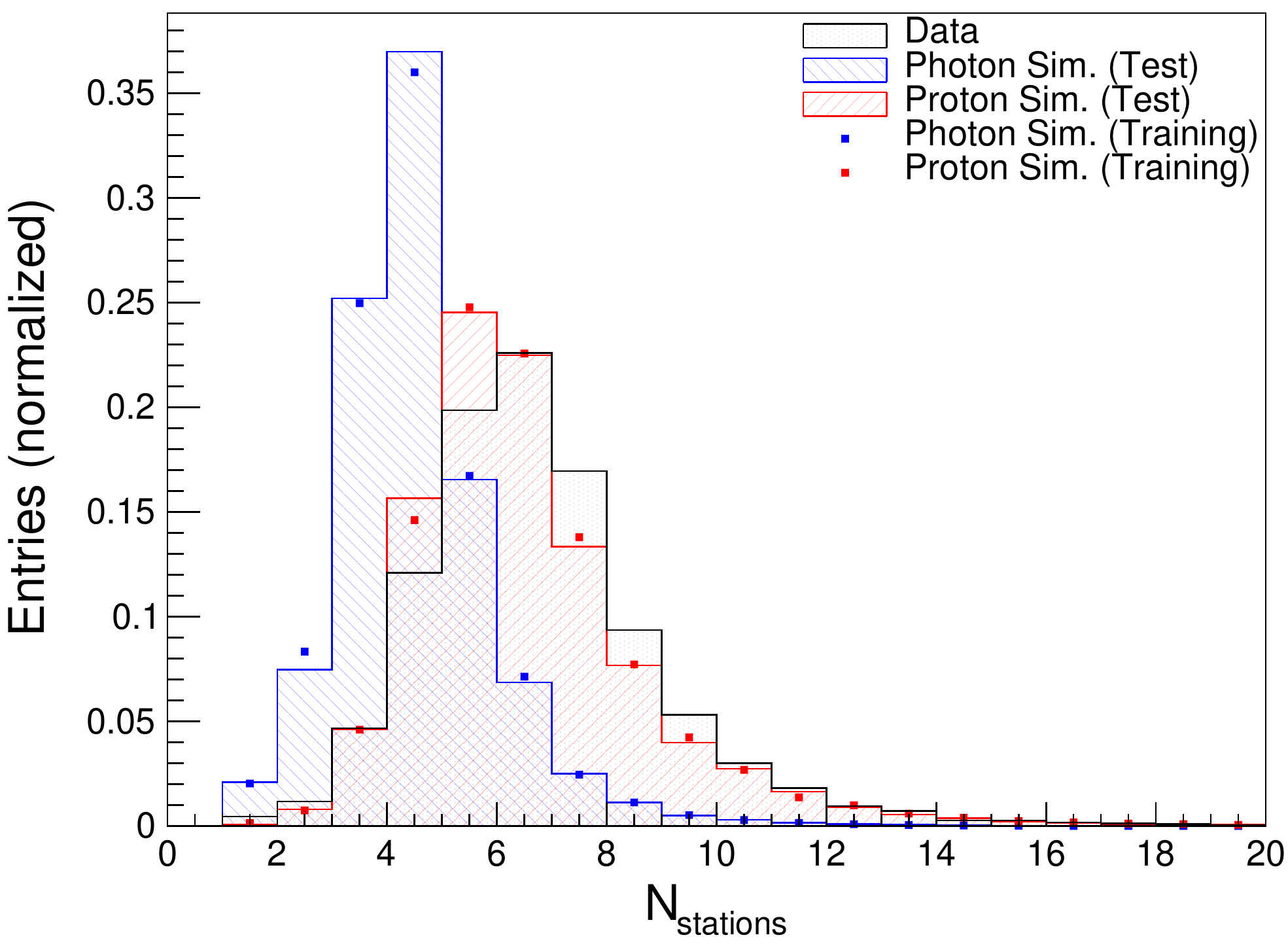}
   \includegraphics[width=0.49\textwidth]{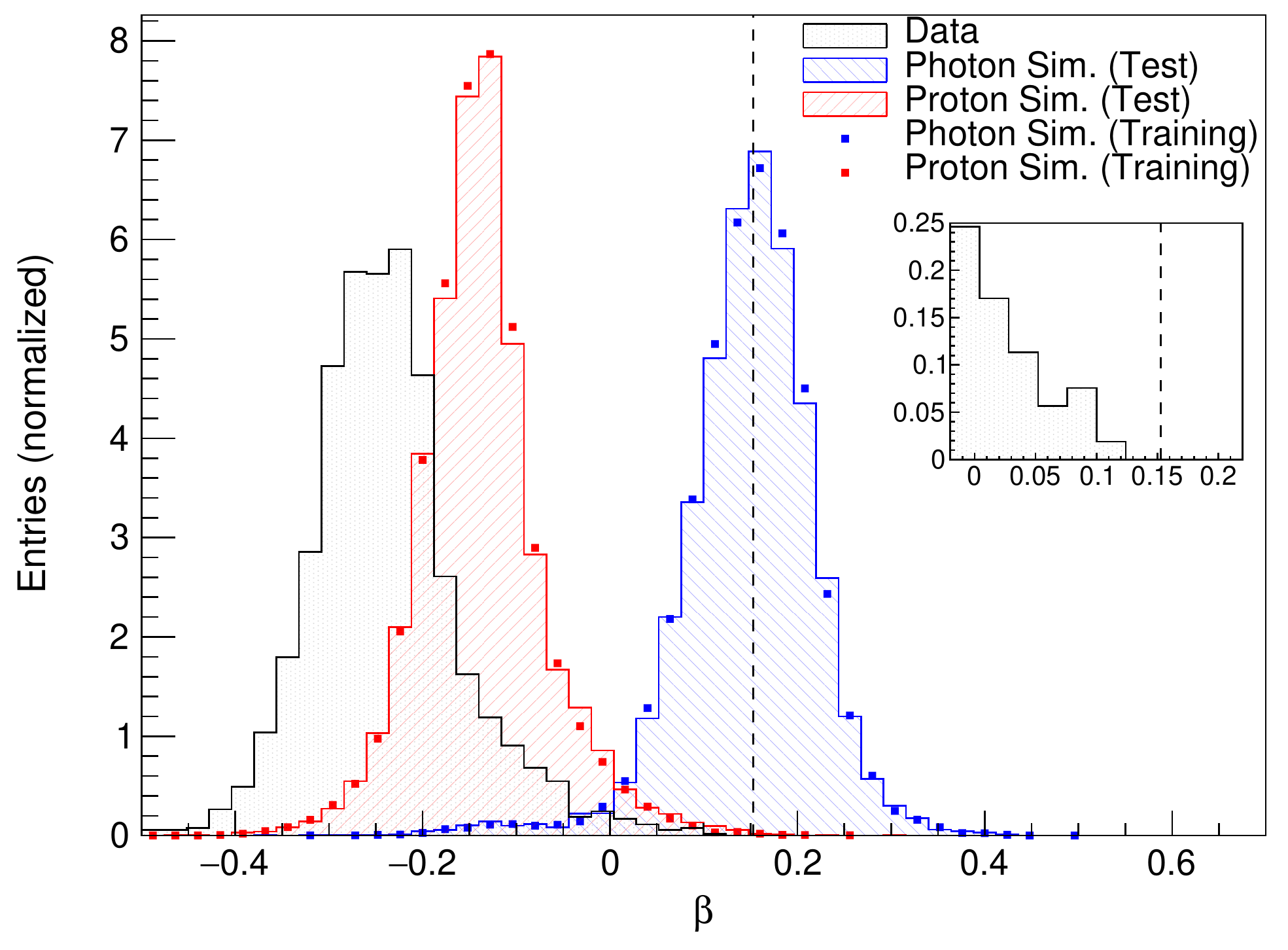}
	\caption{Normalized distributions of the three discriminating observables $X_{\text{max}}$, $S_b$ and $N_{\text{stations}}$ used in the photon search based on HeCo data. The (simulated) photon sample is shown in blue, the (simulated) proton sample in red, and the data sample in black. In addition, normalized distributions of the final discriminator $\beta$---which is based on a MVA combining $X_{\text{max}}$, $S_b$ and $N_{\text{stations}}$ as well as the photon energy and the zenith angle---are displayed. The dashed line denotes the median of the  photon test sample, which is used as the photon candidate cut. In all plots, only events with $E_\gamma{>}\unit[2{\times}10^{17}]{eV}$ are shown. For more details, see~\cite{PierreAuger:2022uwd}.}
	\label{fig:hecoobservables}
\end{figure}

In Fig.~\ref{fig:hecoobservables}, the normalized distributions of the discriminating observables $\xmaxM$, $S_b$ and $N_\text{stations}$ are shown for the simulated samples as well as the data sample. In addition, the corresponding distributions of the output from the BDT $\beta$, which is used as the final discriminator for separating photon-induced air showers from the hadronic background, are displayed. A more detailed discussion of these distributions can be found in~\cite{PierreAuger:2022uwd}. Here, we only note that for $\beta$, the photon and proton distributions are well-separated. The background rejection at a signal efficiency of $\unit[50]{\%}$, i.e., the fraction of proton-induced events with $\beta$ larger than the median of the photon distribution---which is used as the photon candidate cut, marked with the dashed line---is $\unit[(99.91\,{\pm}\,0.03)]{\%}$ for energies $E_\gamma \geq \unit[2{\times}10^{17}]{eV}$.\\

Zero events from the data sample have a $\beta$ value above the candidate cut value, hence no photon candidate events are identified in this analysis. The final results of this study are therefore given in terms of
upper limits on the integral flux of photons
$\Phi^{\text{C.L.}}_{\gamma,\,\text{U.L.}}(E_\gamma{>}E_0)$. The integrated, efficiency-weighted exposure for photons needed to calculate these upper limits is obtained from simulations. In the energy range of interest between $\unit[2{\times}10^{17}]{eV}$ and $\unit[10^{18}]{eV}$, the weighted exposure varies between $2.4$ and $\unit[2.7]{km^2\,sr\,yr}$ under the assumption of a power-law spectrum $\propto E^{-2}$. Upper limits on the integral photon flux are placed at threshold energies of $2$, $3$, and $\unit[5{\times}10^{17}]{eV}$, as well as $\unit[10^{18}]{eV}$, at a confidence level of $\unit[95]{\%}$. At these threshold energies, the upper limits are $2.72$, $2.50$, $2.74$, and $\unit[3.55]{km^{-2}\,sr^{-1}\,yr^{-1}}$, respectively. Using the energy spectrum of cosmic rays measured by the Pierre Auger Observatory~\citep{PierreAuger:2021hun}, the upper limits on the integral photon flux can be translated into upper limits on the integral photon fraction. At a confidence level of $\unit[95]{\%}$, these are $\unit[0.28]{\%}$, $\unit[0.63]{\%}$, $\unit[2.20]{\%}$ and $\unit[13.8]{\%}$ for the same threshold energies as above.


\subsection{A search for ultra-high-energy photons at the Pierre Auger Observatory exploiting air-shower universality}
\label{sec:hybrid}

The photon search in the energy range between $10^{18}$ and $\unit[10^{19}]{eV}$~\cite{PierreAuger:2021mjh} is performed by exploiting the hybrid configuration of the Pierre Auger Observatory, much like in the previous analysis. As before, $\xmaxM$ can be measured directly with the FD. The muon content of a measured air shower is accessed through the parameter $\fmuM$, which is derived from the SD signals by using the air-shower universality concept~\cite{hyb::cit::Lipari2009}. A universality-based model~\cite{AVE201723} is used to predict the signals induced by the secondary particles in the individual SD stations. This model describes the total signal as the sum of four components: two electromagnetic components, one related to high-energy pions ($S_{e\gamma}$) and one related to low-energy hadrons ($S_{e\gamma(\text{had})}$), and two muon-related components, a pure muon component ($S_\mu$) and one related to electrons and photons resulting from muon decays ($S_{e\gamma(\mu)}$). The predicted signal, $S_{\text{pred}}$, can then be expressed as
\begin{equation}
  \label{eq:univ}
  S_{\text{pred}} = \sum_{i=1}^{4} \beta^{i}(\fmuM) \cdot S^i_{\text{comp}} (E, \xmaxM, \text{geometry}),
\end{equation}
where $i$ runs over the four components. Each of the four signal components, $S^i_{\text{comp}}$, has a universal behavior, which can be parameterized as a function of the primary energy $E$, $\xmaxM$, and the geometry of the air shower. The relative contributions $\beta^{i}$ of each of the four components depend only on the mass of the primary particle through a parameter, $\fmuM$,
representing the number of muons in the air shower.\\

\begin{figure}[t!]
  \centering
    \includegraphics[width=0.49\textwidth]{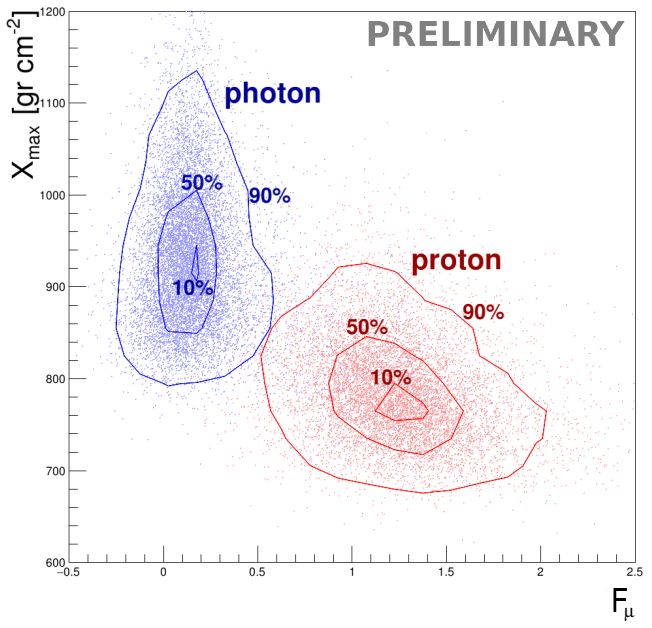}
    \includegraphics[width=0.49\textwidth]{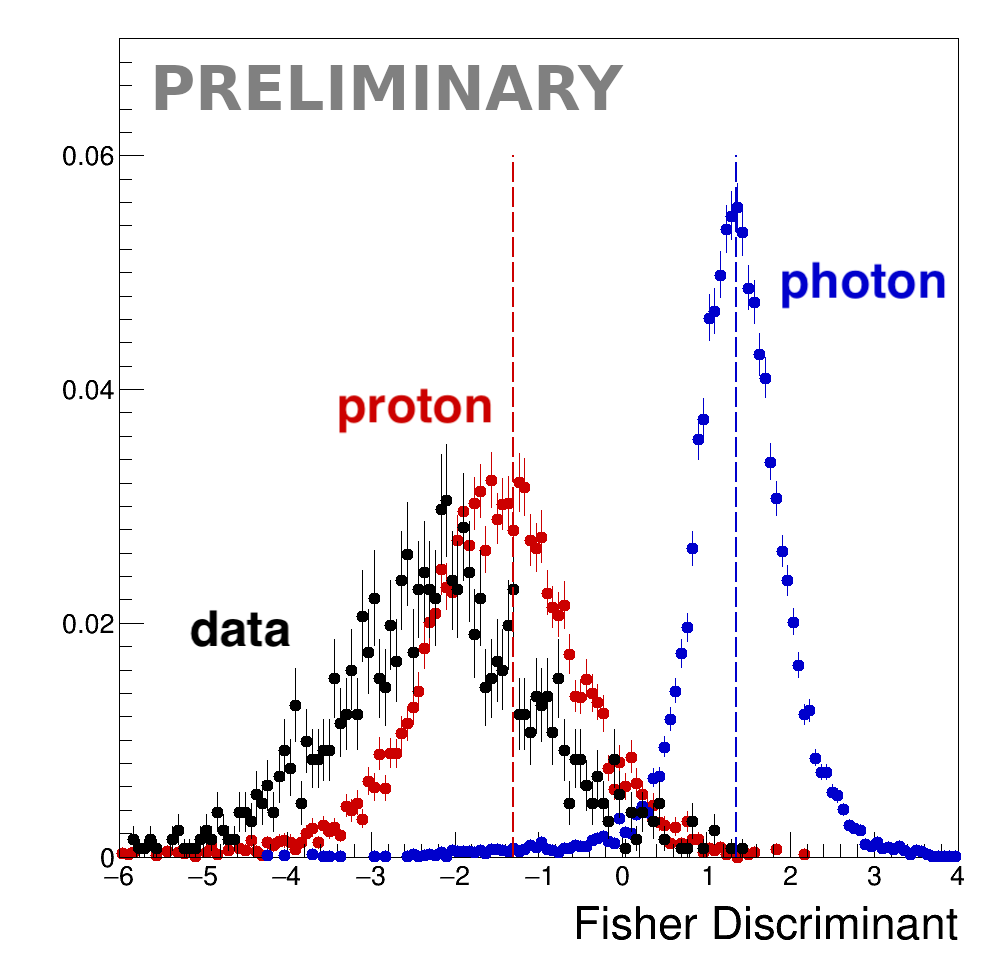}\\
    \includegraphics[width=0.49\textwidth]{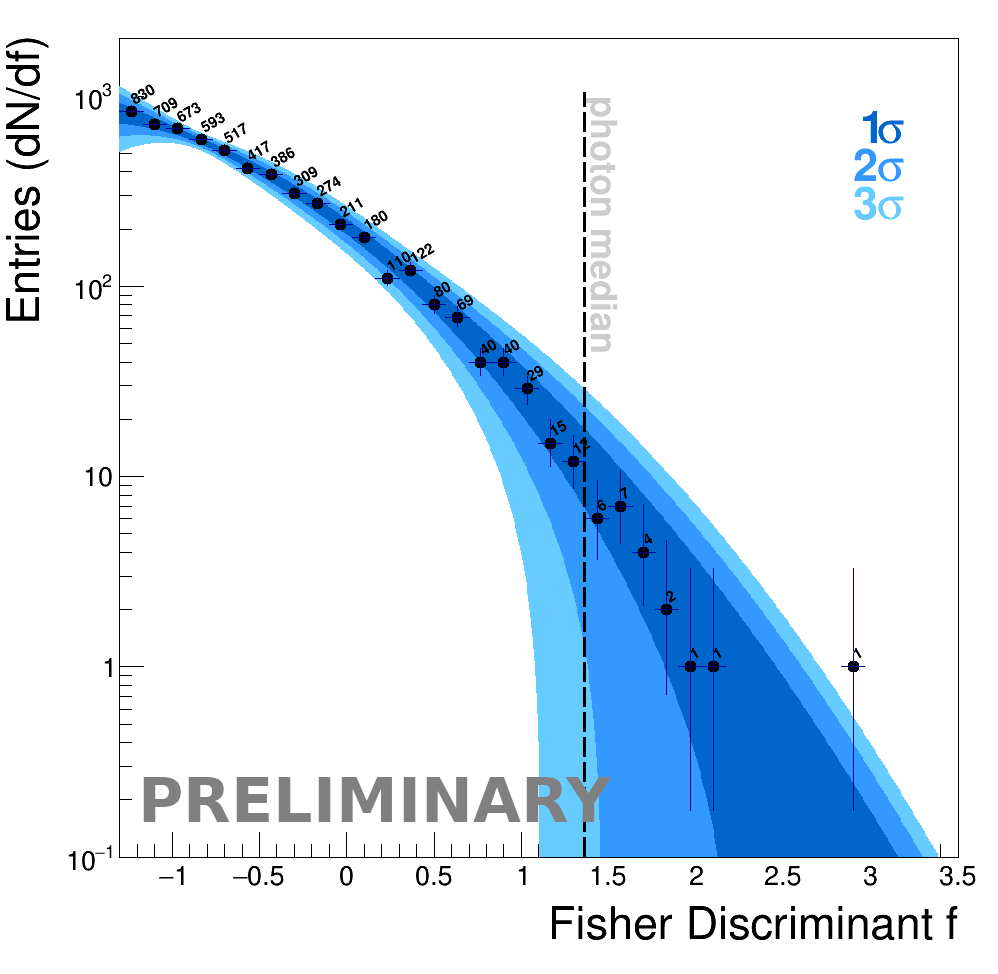} 
  \caption{
    Top left: scatter plot of $\xmaxM$ and $\fmuM$, i.e., the observables used in the hybrid search for photons using air-shower universality, for simulated primary photons (blue) and protons (red); The contour lines enclose $\unit[90]{\%}$, $\unit[50]{\%}$ and $\unit[10]{\%}$, respectively, of the events. Top right: distributions of the Fisher discriminant $f$ for simulated primary photons
 (signal, blue) and protons (background, red), and for the burnt sample (black); the dashed red line marks the tail of the proton distribution, the dashed blue line indicates the median of the photon distribution. Bottom: the tail of the distribution of $f$ for the hybrid data sample (black dots); the dashed line represents the photon-candidate cut; the shaded blue regions show the $1\sigma$, $2\sigma$ and $3\sigma$ uncertainty bands for background expectation. For more details, see~\cite{PierreAuger:2021mjh}.
 }
  \label{hyb::fig::fisher}
\end{figure}

Following the approach developed in~\cite{Savina:2019rtz}, the universality-based signal model is applied to the case of hybrid events measured by the FD and the SD. As the hybrid reconstruction provides $E$, $\xmaxM$ and the shower geometry, the four components $S^i_{\text{comp}}$ can be directly calculated for each SD station involved in a hybrid event. Thus, given the reconstructed signal, $S_{\text{rec}}$, in a station of the SD, $\fmuM$ can be calculated for each station in an event by matching $S_{\text{rec}}$ to $S_\text{pred}$ from Eq.~(\ref{eq:univ}). To obtain an event-wise estimate of the muon content, the average $\fmuM$ of all SD stations is assigned to an event if more than one station is available. Overall, $\fmuM$ provides a very good photon-hadron separation (see Fig.~\ref{hyb::fig::fisher}, top left). To fully exploit the hybrid approach, it is combined in this analysis with $\xmaxM$ in a linear Fisher discriminant analysis~\cite{Fisher:1936et}.\\

The resulting distributions of the Fisher discriminant $f$ are shown in the top right panel of Fig.~\ref{hyb::fig::fisher}, for simulations of primary protons (red) and photons (blue), as well as for $\unit[5]{\%}$ of the data sample (black), which is used as a \textit{burnt sample}. The distributions of $f$ obtained for the proton and photon samples are well separated, resulting in a background rejection of about $\unit[99.90]{\%}$ at a signal efficiency of $\unit[50]{\%}$, corresponding to the dashed blue line in the top right panel of Fig.~\ref{hyb::fig::fisher}. The burnt sample and the photon  distributions are even more separated, therefore the events contained in the burnt sample can be considered as background events only, and can then be used to derive a data-driven estimate of the expected background.  Due to the limited number of events in the burnt sample, we used in a first step the rightmost tail of the proton distribution, specifically only the events with $f > -1.3$ (red dashed line in Fig.~\ref{hyb::fig::fisher}, top right), to derive the functional form of the background distribution. Then, the expected background distribution was derived through a fit to the burnt sample and scaled up linearly with time to match the full data sample.\\

Finally, the analysis has been applied to hybrid events above $\unit[10^{18}]{eV}$ recorded by the Pierre Auger Observatory between 1 January 2005 and 31 December 2017. Of the total data set, which consists of approximately $32{,}000$ events, the rightmost tail of the distribution of the Fisher discriminant $f$ (black dots) is shown in the bottom panel of Fig.~\ref{hyb::fig::fisher}. The data distribution resembles the background expectation, shown as the shaded blue bands, representing the uncertainties in its estimation for different $\sigma$ levels. After applying the photon selection cut (dashed vertical line), corresponding to the median of the photon distribution (dashed blue line in Fig.~\ref{hyb::fig::fisher}, top right), 22 photon-candidate events are selected. This number is consistent with the background expectation of $30 \pm 15$ false-positive candidate events. Since no significant excess with respect to the background has been found,  the final results of this study are given in terms of upper limits on the integral flux of photons
$\Phi^{\text{C.L.}}_{\gamma,\,\text{U.L.}}(E_\gamma{>}E_0)$.  
Five different energy thresholds ($1$, $2$, $3$, and $\unit[5{\times}10^{18}]{eV}$, as well as $\unit[10^{19}]{eV}$) were considered. The number of photon-candidate events found for each energy threshold were 22, 2, 0, 0 and 0, respectively. The upper limits are determined taking into account the expected number of background events derived from the burnt sample for each threshold ($30 \pm 15$, $6 \pm 6$, $0.7 \pm 1.9$, $0.06 \pm 0.25$ and $0.02 \pm 0.06$, respectively) as well as the integrated, efficiency-weighted exposure for photons, which was again determined from simulations. In the energy range between $10^{18}$ and $\unit[10^{19}]{eV}$, the weighted exposure increases from $420.7$ to $\unit[1245.9]{km^{2}\,sr\,yr}$, under the assumption of a power-law spectrum $\propto E^{-2}$. The resulting upper limits (at a confidence level of $\unit[95]{\%}$) on the integral flux of photons for the aforementioned thresholds are $4.0$, $1.1$, $0.35$, $0.23$ and $\unit[0.0021]{km^{-2}\,sr^{-1}\,yr^{-1}}$. These results are preliminary and will be updated in a forthcoming journal publication.


\subsection{Search for photons above \texorpdfstring{$\unit[10^{19}]{eV}$}{1019 eV} with the surface detector of the Pierre Auger Observatory}
\label{sec:sd}

In the energy range above $\unit[10^{19}]{eV}$, UHE photons are searched for among the data collected with the $\unit[1500]{m}$ SD array of the Pierre Auger Observatory~\cite{PierreAuger:2022abc}. While the photon search using SD-only data can profit from the large exposure due to the high duty cycle of the SD, the lack of a corresponding fluorescence measurement for the bulk of the data poses some challenges. For example, there is no direct measurement of $\xmaxM$ available. Also the primary energy can only be accessed indirectly, using $S(1000)$---the interpolated signal in the SD stations at a perpendicular distance of $\unit[1000]{m}$ from the shower axis---as a proxy.\\

\begin{figure}[t!]
   \centering
   \includegraphics[width=0.49\textwidth]{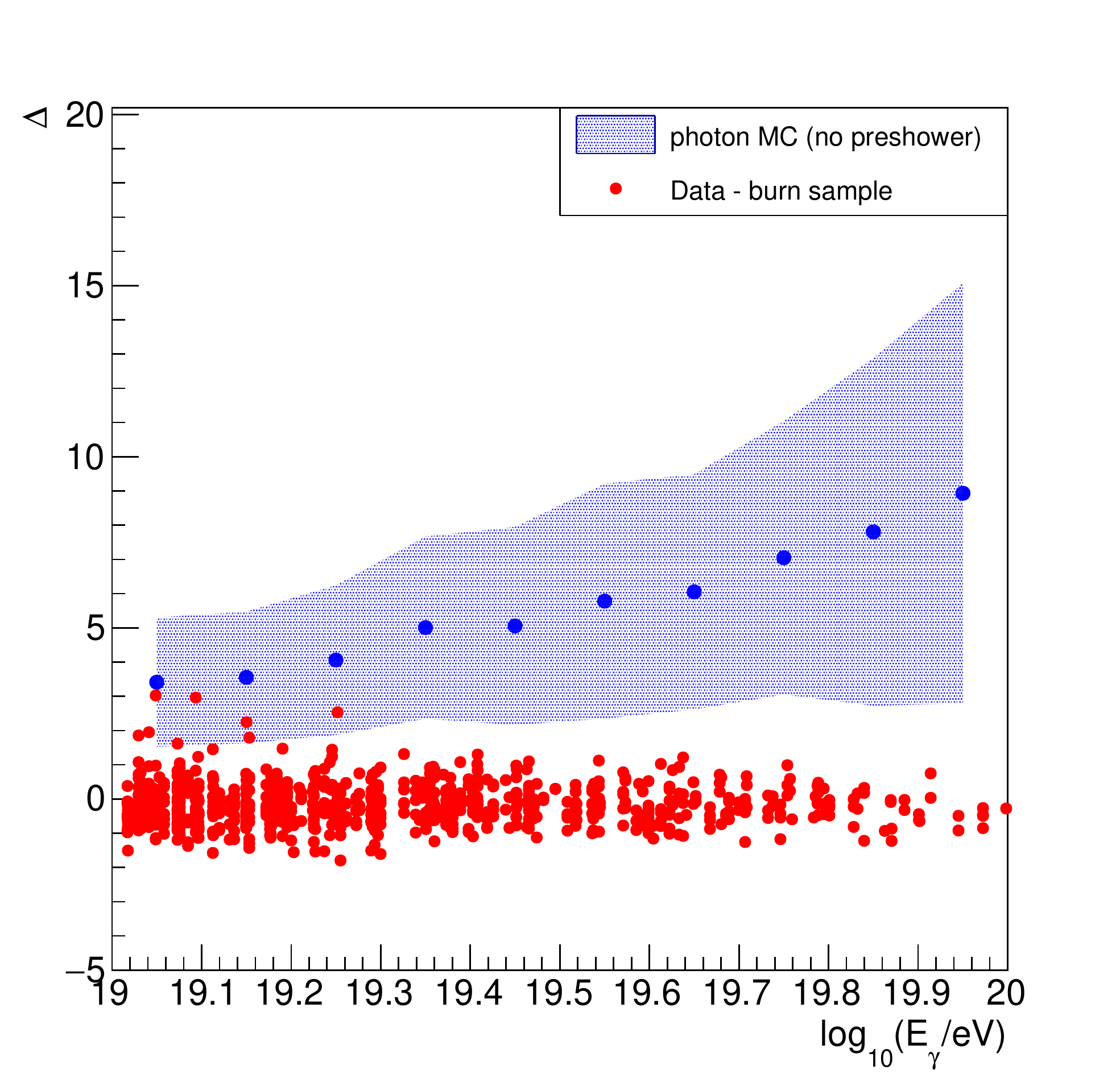}
   \includegraphics[width=0.49\textwidth]{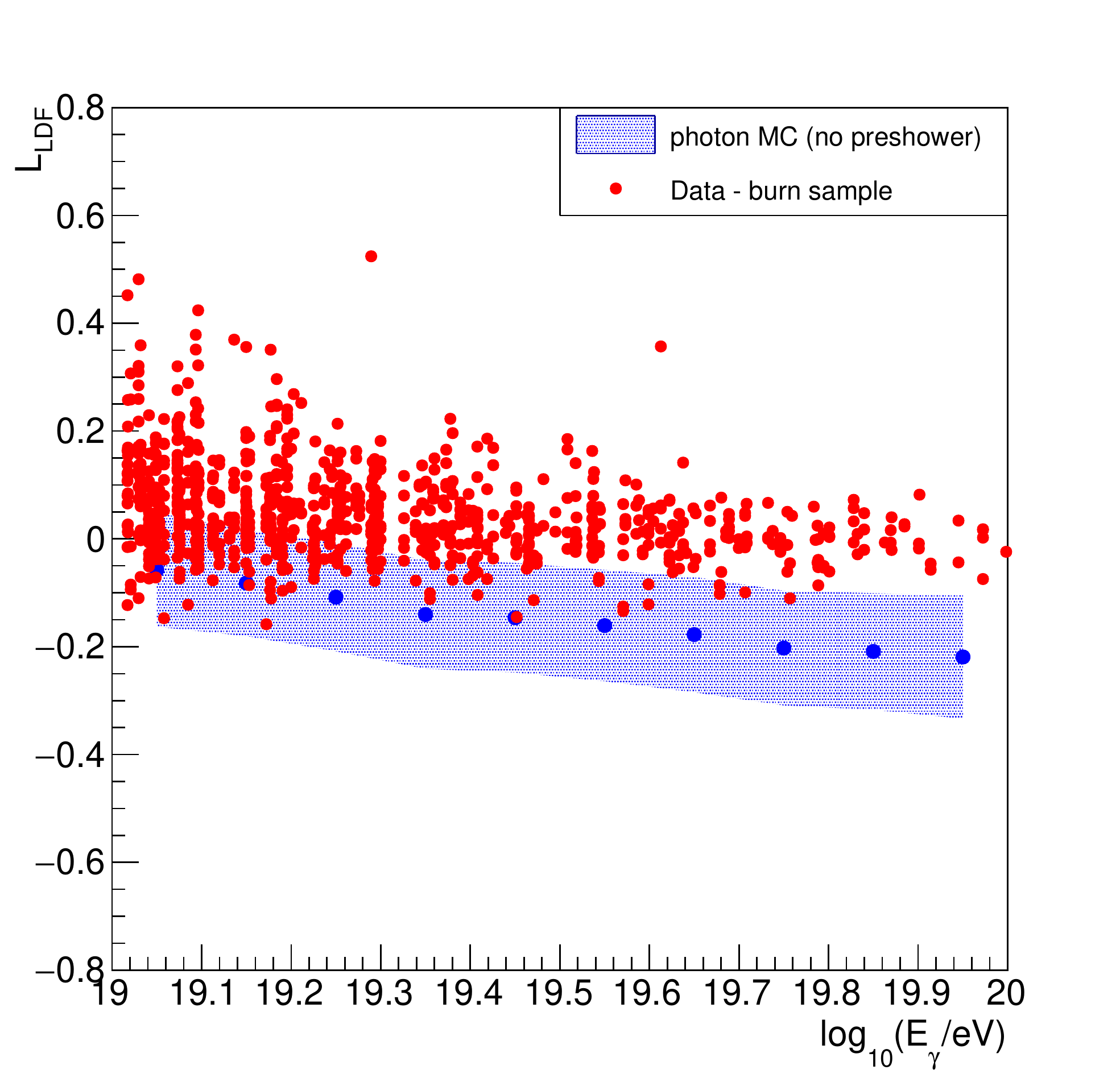}
   \includegraphics[width=0.49\textwidth]{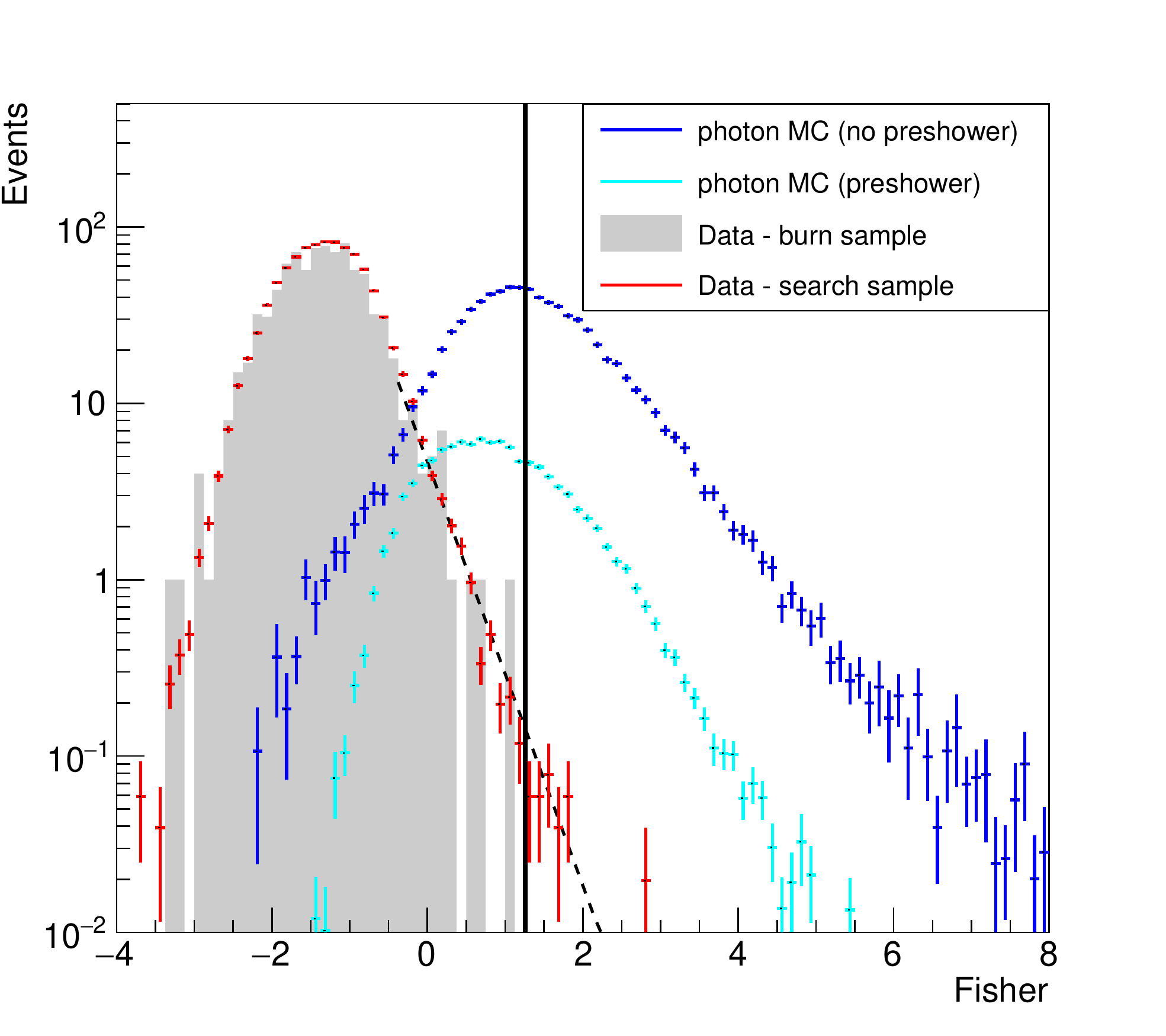}
 	\caption{Top: distributions of $\Delta$ (left) and $L_{\rm LDF}$ (right), i.e., the observables used in the search for photons based on SD-only data, as a function of the photon energy $E_\gamma$ for simulated primary photons (in blue) and a fraction of the data sample (in red) that is used as a burnt sample; the bands represent one standard deviation of the photon
distributions. Bottom: distributions of the Fisher discriminant for the burnt sample (grey), the search sample (red) and simulated primary photons (non-preshowering in blue and preshowering in light blue), weighted with an $E^{-2}$ spectrum; the search sample and the photon distributions are scaled as to have the same integral as the burn sample one; the vertical line indicates the value of the photon-candidate cut; the dashed line shows the result of the fit of an exponential to the 5\% of events in the burnt sample with the largest values of the Fisher discriminant. For more details, see~\cite{PierreAuger:2022abc}.}
	\label{fig:sdsearch}
\end{figure}

Two observables are used in this analysis, one related to the thickness of the shower front at ground and one based on the steepness of the lateral distribution. These two properties of an air shower depend on the type of the primary particle initiating the shower, hence they can be used for photon-hadron separation. The first observable, $\Delta$, is based on the risetime $t_{1\mspace{-2mu}/\mspace{-2mu}2}$ in the individual SD stations, which is defined as the time at which the integrated signal in the measured time trace rises from $\unit[10]{\%}$ to $\unit[50]{\%}$ of its total value. For showers of the same primary energy and zenith angle, $t_{1\mspace{-2mu}/\mspace{-2mu}2}$ is expected to be larger for primary photons with respect to primary nuclei due to the reduced muonic content, which implies larger scattering and attenuation of secondary particles, and to $\xmaxM$ being closer to the ground. $\Delta$ is defined as: 
\begin{equation}
    \Delta = \frac{1}{N} \sum_i \frac{(t^i_{1\mspace{-2mu}/\mspace{-2mu}2}-t^\mathrm{bench}_{1\mspace{-2mu}/\mspace{-2mu}2})}{\sigma_{t_{1\mspace{-2mu}/\mspace{-2mu}2}}},
\end{equation}
which can be taken as the average deviation of the measured rise-times from a \textit{data benchmark} $t_{1\mspace{-2mu}/\mspace{-2mu}2}^\mathrm{bench}$, describing the average risetime of all of the SD data (assumed to be overwhelmingly constituted by primary nuclei)~\cite{PierreAuger:2017tlx}, in units of sampling fluctuations $\sigma_{t_{1\mspace{-2mu}/\mspace{-2mu}2}}$. Details on the selection criteria for the SD stations can be found in~\cite{PierreAuger:2022abc}. By construction, $\Delta$ is expected to average to zero for data and to be significantly positive for photon-induced air showers. As photon-induced air showers are also expected to have a steeper lateral distribution of the signals in the SD stations than the average of all SD data, a second observable $L_{\rm LDF}$ is introduced to quantify the departure of the observed lateral distribution function (LDF) from the average of all SD data (see also \cite{PierreAuger:2020yab}):
\begin{equation}
    L_{\rm LDF} = \log_{10}\left( \frac{1}{N} \sum_{i=1}^N \frac{S_i}{f_{\rm LDF}(r_i)}\right),
\end{equation}
where $S_i$ is the total signal measured in the $i$-th selected station and $f_{\rm LDF}(r_i)$ gives the average signal, obtained from all SD data, for a station at the same distance $r_i$ from the shower axis. The photon energy $E_\gamma$ is determined for each measured air-shower event taking into account $S(1000)$ and the reconstructed zenith angle. For this purpose, a look-up table has been constructed using a large simulation sample. Only non-preshowering photon events are used, which are weighted according to a reference spectrum $\propto E^{-2}$. In Fig.~\ref{fig:sdsearch}, the distributions of the two observables are shown as a function of the photon energy. In particular $\Delta$ shows a good separation between photons and data. Finally, the two variables are combined using a Fisher discriminant analysis with the burnt sample representing the background and photon simulations the signal.\\

The analysis is applied to SD data collected between 1 January 2004 and 30 June 2020. Only air-shower events with a reconstructed zenith angle between $30^\circ$ and $60^\circ$ are taken into account to ensure that the majority of possible selected photon-induced showers reach their maximum development before reaching ground level. A number of selection criteria are applied to ensure a reliable reconstruction of the two observables. These criteria are described in detail in~\cite{PierreAuger:2022abc}. Overall, the data sample (search sample) consists of $48{,}061$ selected events with a photon energy $E_\gamma \geq \unit[10^{19}]{eV}$, excluding the burnt sample which consists of $886$ events ($\unit[1.8]{\%}$ of the total number of selected events). The results of the analysis are shown in Fig.~\ref{fig:sdsearch}, bottom. 16, 2 and 0 events above energy thresholds of $1$, $2$ and $\unit[4{\times}10^{19}]{eV}$, respectively, had a value of the Fisher discriminant above the photon-candidate cut, which was fixed to the median of the Fisher distribution for non-preshowering primary photons (shown as the solid black line in Fig.~\ref{fig:sdsearch}, bottom). The number of observed candidate events is in statistical agreement with what is expected from the fit of an exponential to the tail of the distribution of the Fisher discriminant for the burnt sample. In addition, no peak-like features, which would indicate the presence of a photon population, are observed above the fall-off of the distribution. Overall, the results are consistent with the expectation for a background of UHE protons and nuclei, hence upper limits on the integral flux of UHE photons are determined. To calculate these upper limits, the signal efficiency of the analysis is required. The efficiency has been determined from simulations, and it increases from $0.26$ for a threshold energy of $\unit[10^{19}]{eV}$ to $0.39$ for $\unit[4{\times}10^{19}]{eV}$, under the assumption of a power-law spectrum $\propto E^{-2}$. Upper limits on the integral photon flux are placed at threshold energies of $1$, $2$, and $\unit[4{\times}10^{19}]{eV}$, at a confidence level of $\unit[95]{\%}$. At these threshold energies, the upper limits are $2.11$, $0.312$, and $\unit[0.172{\times10^{-3}}]{km^{-2}\,sr^{-1}\,yr^{-1}}$, respectively. These upper limits on the integral flux of photons correspond to upper limits on the integral photon fraction of $\unit[1.6]{\%}$, $\unit[1.2]{\%}$, and $\unit[3.2]{\%}$, for the same threshold energies and again for a confidence level of $\unit[95]{\%}$. As before, the fraction limits have been calculated using the most up-to-date measurement of the energy spectrum of UHE cosmic rays from the Pierre Auger Observatory~\citep{PierreAuger:2021hun}. 


\subsection{Summary of the searches for a diffuse flux of UHE photons}
\label{sec:diffusesummary}

\begin{table}[t]
  \begin{center}
    \begin{tabular}{c|c|c|c}
    \hline \hline
    Detector &
     $\unit[E_0]{[eV]}$ & $\unit[\Phi^{\unit[95]{\%}}_{\gamma,\,\text{U.L.}}(E_\gamma {>}E_0)]{[km^{-2}\,yr^{-1}\,sr^{-1}]}$ & Reference \\ 
    \hline
    \multirow{4}{*}{\rotatebox[origin=c]{90}{\parbox[c]{50pt}{\centering HeCo + \\SD $\unit[750]{m}$}}} & $2{\times}10^{17}$ & $2.72$ & \multirow{4}{*}{\cite{PierreAuger:2022uwd}}\\
     & $3{\times}10^{17}$ & $2.50$ & \\
     & $5{\times}10^{17}$ & $2.74$ & \\
     & $10^{18}$ & $3.55$ & \\
    \hline
     \multirow{5}{*}{\rotatebox[origin=c]{90}{\parbox[c]{50pt}{\centering FD + \\SD $\unit[1500]{m}$}}} & $10^{18}$ & $4{\times}10^{-2}$ & \multirow{5}{*}{\cite{PierreAuger:2021mjh}}\\
     & $2{\times}10^{18}$ & $1.1{\times}10^{-2}$ & \\
     & $3{\times}10^{18}$ & $0.35{\times}10^{-2}$ & \\
     & $5{\times}10^{18}$ & $0.23{\times}10^{-2}$ & \\
     & $10^{19}$ & $0.21{\times}10^{-2}$ & \\    
    \hline
     \multirow{3}{*}{\rotatebox[origin=c]{90}{\parbox[c]{50pt}{\centering SD $\unit[1500]{m}$}}} & $10^{19}$ & $2.11{\times}10^{-3}$ & \multirow{3}{*}{\cite{PierreAuger:2022abc}}\\
     & $2{\times}10^{19}$ & $0.312{\times}10^{-3}$ & \\
     & $4{\times}10^{19}$ & $0.172{\times}10^{-3}$ & \\
    \hline \hline
    \end{tabular}
  \caption{Compilation of the upper limits on the integral photon flux determined through the three analyses discussed in the previous sections.}
  \label{tab:alllimits}
  \end{center}
\end{table}

\begin{figure}[t]
   \centering
   \includegraphics[width=0.9\textwidth]{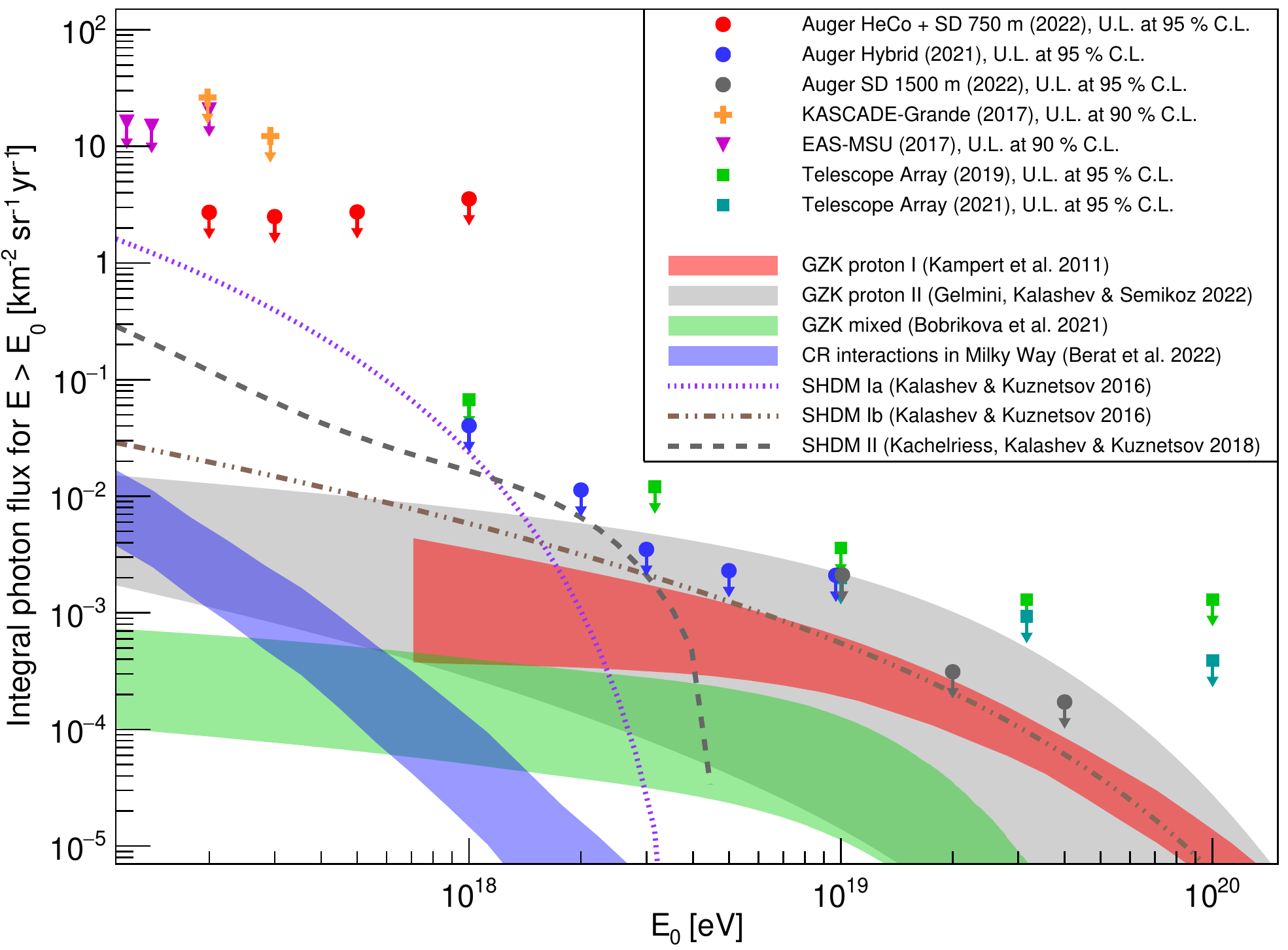}
\caption{
Current upper limits on the integral photon flux determined from data collected by the Pierre Auger Observatory (red, blue and gray circles). Shown are also upper limits published by other experiments: KASCADE-Grande (orange crosses)~\cite{Apel:2017ocm}, EAS-MSU (magenta triangles)~\cite{Fomin:2017ypo}) and Telescope Array (green squares from~\cite{Abbasi:2018ywn} and turquoise squares from~\cite{TA:2021mko}). The ranges of expected GZK photon fluxes under the assumption of two different pure-proton scenario are shown as the red and gray bands (following~\cite{Kampert:2011hkm} and~\cite{Gelmini:2022sti}, respectively). The green band shows the expected GZK photon flux assuming a mixed composition that would fit the Auger data~\cite{Bobrikova:2021kuj}, while the blue band denotes the range of photon fluxes that would be expected from cosmic-ray interactions with matter in the Milky Way~\cite{Berat2021}. In addition, the expected photon fluxes from the decay of super-heavy dark matter particles are included (decay into hadrons, $X\rightarrow q\bar{q}$, based on~\cite{Kalashev:2016cre}: dashed violet line for a mass of the SHDM particles $M_X=\unit[10^{10}]{GeV}$ and a lifetime $\tau_X=\unit[3{\times}10^{21}]{yr}$ [SHDM Ia]; brown dot-dashed line for $M_X=\unit[10^{12}]{GeV}$ and $\tau_X=\unit[10^{23}]{yr}$ [SHDM Ib]; decay into leptons, $X\rightarrow \nu\bar{\nu}$, based on~\cite{Kachelriess:2018rty}: dashed gray line for $M_X=\unit[10^{10}]{GeV}$  and $\tau_X=\unit[3{\times}10^{21}]{yr}$ [SHDM II]; the exact lines have been obtained through personal communication with one of the authors).
}
\label{fig:UL}
\end{figure}

The upper limits on the integral photon flux derived through the three analyses discussed in the previous sections are compiled in Tab.~\ref{tab:alllimits} and shown in Fig.~\ref{fig:UL}, together with upper limits published by other experiments. The Pierre Auger Observatory currently provides the most stringent limits over a wide energy range, spanning from $\unit[2{\times}10^{17}]{eV}$ to the highest energies. In addition, the set of upper limits derived from HeCo data closed the gap between the upper limits at ultra-high energies (derived from hybrid and SD data) and those determined by smaller air-shower experiments such as KASCADE-Grande, leading to a full coverage of the aforementioned energy range. It is worth mentioning here that extensive systematic studies have been performed to test the robustness of the analyses and their results against various sources of uncertainties, for example in the hadronic interaction models used in the air-shower simulations or in the reconstruction of the different observables. Overall, the results proved to be very robust, more details on these studies can be found in~\cite{PierreAuger:2022uwd,PierreAuger:2021mjh,PierreAuger:2022abc}.\\

For comparison, also the expected fluxes of UHE photons under different theoretical assumptions are shown in Fig.~\ref{fig:UL}. First, we discuss briefly the expected fluxes resulting from interactions of UHECRs with the background photon fields permeating the Universe, most notably the cosmic microwave background~\cite{Greisen:1966jv,Zatsepin:1966jv}. In Fig.~\ref{fig:UL}, the expectations for two different pure-proton scenarios~\cite{Kampert:2011hkm, Gelmini:2022sti} are shown, as well as a scenario involving a mixed composition at the sources~\cite{Bobrikova:2021kuj}. While the experimental sensitivities reached above $\approx\unit[3{\times}10^{18}]{eV}$ start to approach or already constrain the most optimistic expectations of the cosmogenic photon flux from protons, they are about $1$ to $1.5$ orders of magnitude above those from the mixed-composition model. Another cosmogenic flux is that from the interactions of UHECRs with the matter traversed in the Galactic plane~\cite{Berat2021}. While this flux becomes comparable to the aforementioned ones below $\unit[10^{18}]{eV}$, they are still two to three orders of magnitude below the current upper limits in this energy region. Finally, UHE photons could also result from the decay of super-heavy dark matter (SHDM) particles. It should be noted that previous upper limits on the incoming photon flux already severely constrained non-acceleration models in general, and SHDM models in particular, trying to explain the origin of cosmic rays at the highest energies (see, e.g.,~\cite{PierreAuger:2007hjd,Aab:2016agp}). With the upper limits on the incoming photon flux, it is possible to constrain the phase space of mass and lifetime of the SHDM particles~\cite{Anchordoqui:2021crl}. As an example, we show the expectations for three different assumptions: For a hadronic decay ($X \rightarrow q\bar{q}$), we show the expected fluxes according to~\cite{Kalashev:2016cre} for a mass $M_X$ of the SHDM particles of $\unit[10^{10}]{GeV}$ and a lifetime $\tau_X$ of $\unit[3{\times}10^{21}]{yr}$, as well as for $M_X = \unit[10^{12}]{GeV}$ and a lifetime $\tau_X = \unit[10^{23}]{yr}$. Both combinations are currently allowed. Since also a decay into leptons ($X\rightarrow \nu\bar{\nu}$) is possible, we show the expected flux according to~\cite{Kachelriess:2018rty} for $M_X = \unit[10^{10}]{GeV}$ and a lifetime $\tau_X = \unit[3{\times}10^{21}]{yr}$. As the sensitivity of current photon searches increases, it will be possible to further constrain these values~\cite{Anchordoqui:2021crl}.


\section{Searches for UHE photons from point sources and transient events}
\label{sec:searchespointsources}

The analyses discussed in the previous section are searches for a diffuse, i.e., direction-independent flux of UHE photons. Naturally, the arrival direction of a cosmic particle carries important information. Photons, like neutrinos, are neutral particles. They are therefore not deflected by galactic and extragalactic magnetic fields and point right back at their production site. This can be used to search for point sources of UHE photons by simply looking for an excess of events from a certain direction~\cite{PierreAuger:2014eyz}, taking into account the angular resolution of the experiment. Complementary to a blind search over the full visible sky, the search for photons can also be restricted to the directions of putative sources to reduce the statistical penalty~\cite{PierreAuger:2016ppv}. In the following sections, we briefly summarize these two analyses. In addition, we discuss follow-up searches for UHE photons from transient events, for example in association with gravitational-wave events~\cite{ruehl:2021rmn}.


\subsection{A search for point sources of EeV photons}
\label{sec:blind}

\begin{figure}[t]
   \centering
   \includegraphics[width=0.75\textwidth]{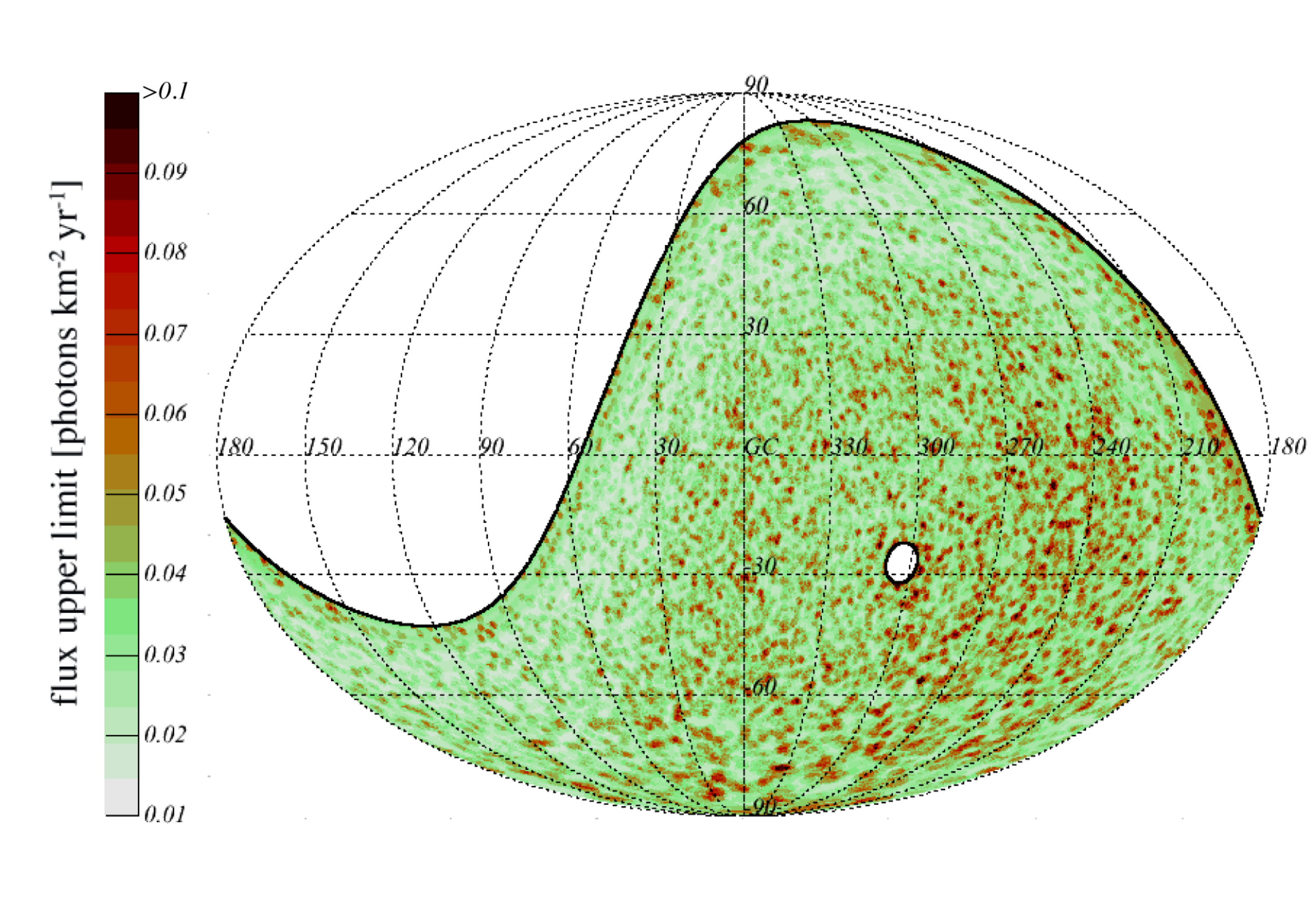}
\caption{
Celestial map, in Galactic coordinates, of upper limits on the incoming photon flux~\cite{PierreAuger:2014eyz}. The white regions indicate regions of the sky that are either not in the field of view of the Pierre Auger Observatory (northern hemisphere) or omitted in this analysis (southern celestial pole). For more details, see~\cite{PierreAuger:2014eyz}.
}
\label{fig:blindsearch}
\end{figure}

The blind search for point sources of UHE photons is performed in the energy range between $\unit[10^{17.3}]{eV}$ and $\unit[10^{18.5}]{eV}$ using hybrid data collected between January 2005 and September 2011. The data set covers a declination range between $-85^\circ$ and $+20^\circ$. The average angular resolution of this data
set is $0.7^\circ$. To reduce the contamination of (isotropically distributed) hadronic background events, 
photon-like air showers are selected using a BDT. The main input
variables of the BDT are $X_\text{max}$ and $S_b$, complemented by additional
observables based on the fit of a Greisen function to the recorded longitudinal profile, and the ratio
of the early-arriving to the late-arriving signal in the SD station with the highest signal~\cite{PierreAuger:2014eyz}. The selection of photon-like events is optimized for
each direction by taking into account the expected number of
background events, which has been derived using the
scrambling technique~\cite{CASSIDAY1990291}. No evidence for an excess of photon-like
events has been found for any direction within the declination band specified before. The resulting upper limits on the flux of UHE photons for each direction are shown in Fig.~\ref{fig:blindsearch}. The average upper limit on the particle flux is $\unit[0.035]{km^{-2}\,yr^{-1}}$ with a maximum of $\unit[0.14]{km^{-2}\,yr^{-1}}$, corresponding to upper limits on the energy flux of $\unit[0.06]{eV\,cm^{-2}\,s^{-1}}$ (average) and $\unit[0.25]{eV\,cm^{-2}\,s^{-1}}$ (maximum), under the assumption of an energy spectrum following a power law with spectral index $-2$.


\subsection{A targeted search for point sources of EeV photons with the Pierre Auger Observatory}
\label{sec:targeted}

\begin{table}[t]
  \begin{center}
    \begin{tabular}{l|ccc||ccc}
    \hline \hline
     Class & $N$ & $\mathcal{P}$ & $\mathcal{P}_w$  & $p$ & $p^*$ & $f_\text{UL}^{0.95}$ [km$^{-2}$~yr$^{-1}$] \\ 
    \hline
      msec pulsars                & 67  &  0.14 & 0.57 & 0.010 & 0.476 & 0.043\\      
     $\gamma$-ray pulsars & 75  & 0.98 & 0.97 & 0.007 & 0.431 & 0.045 \\      
     Low-mass X-ray binaries                         & 87 & 0.74 & 0.13 & 0.014 & 0.718 & 0.046  \\      
     High-mass X-ray binaries                        & 48 & 0.84 & 0.33 & 0.040 & 0.856 & 0.036   \\     
     H.E.S.S. pulsar wind nebulae             & 17 & 0.90 & 0.92 & 0.104 & 0.845 & 0.038     \\          
     H.E.S.S. other            & 16 & 0.52 & 0.12 & 0.042 &  0.493 & 0.040   \\          
     H.E.S.S. unidentified            & 20 &  0.45 & 0.79 & 0.014 & 0.251 & 0.045  \\      
     Microquasars            & 13 &  0.48 & 0.29 & 0.037 &   0.391 & 0.045  \\          
     Magnetars                 & 16 & 0.89 & 0.30 & 0.115 & 0.858  & 0.031   \\      
     Galactic Center              & 1 & 0.59 & 0.59 & 0.471 & 0.471 & 0.024  \\     
     Large Magellanic Cloud                          & 3 & 0.62 & 0.52 & 0.463 & 0.845 & 0.030   \\       
     Centaurus A                       & 1 & 0.31 & 0.31 & 0.221 & 0.221 & 0.031 \\ \hline \hline
     \end{tabular}
    \caption{Combined unweighted probabilities $\mathcal{P}$ and
      weighted probabilities $\mathcal{P}_w$ for the 12 target
      sets analyzed in~\cite{PierreAuger:2016ppv}. In addition, selected information on the
      most significant target from each target set is given: the
      unpenalized ($p$) and penalized ($p^*$) $p$-values and the derived upper
      limit on the photon flux at $\unit[95]{\%}$ C.L.. More details
      on the most significant targets, e.g. the galactic coordinates
      and upper limits on the energy flux, can be
      found in~\cite{PierreAuger:2016ppv}.}
  \label{tab:targetedresults}
  \end{center}
\end{table}

The targeted search for point sources of UHE photons follows the same analysis logic as the blind search summarized before, albeit with a larger data set (January 2005 to December 2013). To reduce the statistical penalty of looking at all directions in the visible sky, the targeted search is restricted to 12 pre-defined target classes, containing in total 364 targets. Since the attenuation length of
photons in the energy range considered here ($10^{17.3}$ to $\unit[10^{18.5}]{eV}$, same as before) varies between $90$ and
$\unit[900]{kpc}$~\cite{PierreAuger:2016ppv}, these target classes contain mostly
galactic sources such as, e.g., millisecond pulsars, $\gamma$-ray pulsars,
and low-mass and high-mass X-ray binaries as well as the Galactic center. In addition, two
nearby extragalactic target sets are included: three powerful
$\gamma$-ray emitters in the Large Magellanic Cloud and the core
region of Centaurus A. The different target classes are listed in
Tab.~\ref{tab:targetedresults}.\\

A $p$-value $p_i$ is assigned to each candidate source $i$ of a target
set, taking into account the observed number of events from this
target direction as well as the expected number of background
events. The $p$-values of all targets in a set are combined with and
without statistical weights, which take into account both the measured electromagnetic flux from the source
(taken from astrophysical catalogs) and the directional exposure for
photons, derived from simulations. The combined weighted probability $\mathcal{P}_{w}$ is the fraction of isotropic simulations yielding a weighted product that is not greater than the measured weighted product. The combined unweighted probability
$\mathcal{P}$ is calculated similarly, but with equal weights for all
targets. The results of the analysis for
each of the 12 target sets are shown in
Tab.~\ref{tab:targetedresults}, along with information about the
target with the smallest $p$-value in each set. In addition, the
penalized $p$-values $p^*$ are given, i.e., the chance
probability that one or more of the targets in the set have
a $p$-value less than $p$ under the assumption of a uniform
probability distribution. No combined $p$-value (weighted and
unweighted) nor any individual $p$-value for a target has a
statistical significance as great as $3\sigma$. No target class
therefore reveals compelling evidence for photon-emitting sources in
the $\unit[10^{18}]{eV}$ range. There is also no evidence for one outstanding
target in any target set.


\subsection{Follow-up search for UHE photons from gravitational wave sources with the Pierre Auger Observatory}
\label{sec:gwfollowup}

\begin{figure}[t]
   \centering
   \includegraphics[width=0.49\textwidth]{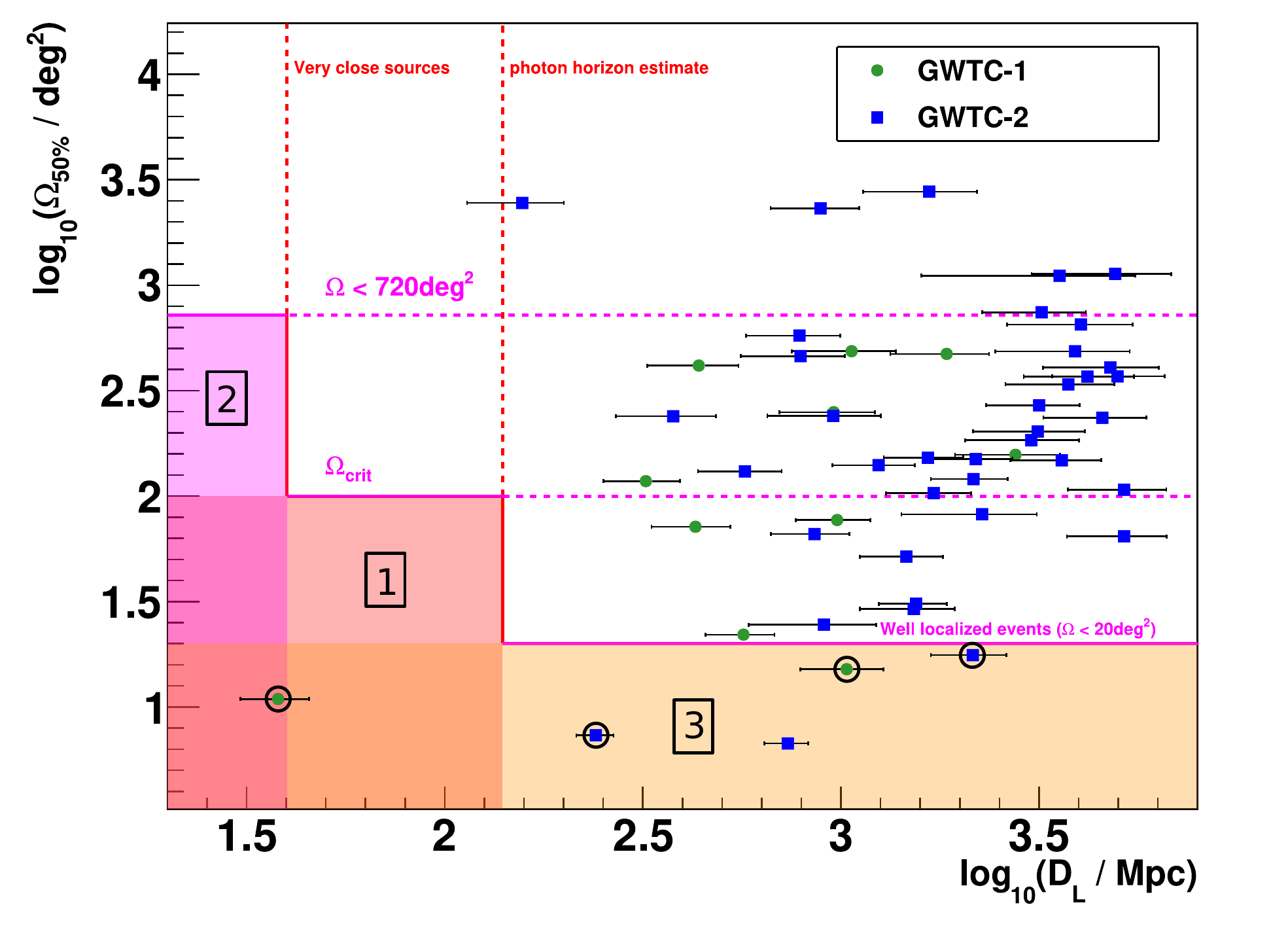}
   \includegraphics[width=0.49\textwidth]{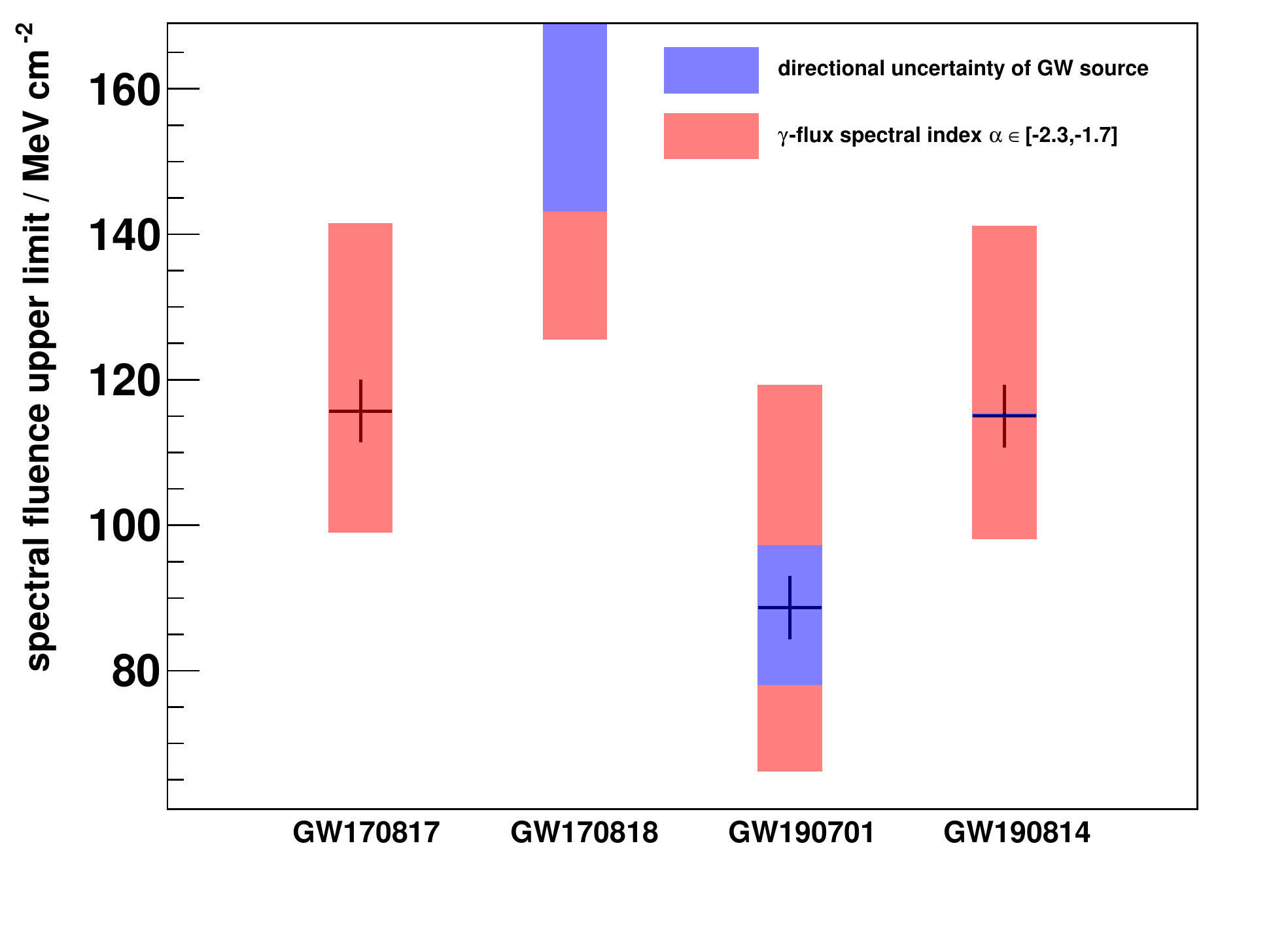}
\caption{
Left: The three classes of selected gravitational wave sources in the follow-up search for photons in association with gravitational wave events, as defined by their $\unit[50]{\%}$ localization region ($\Omega_{50\%}$) and luminosity distance ($D_L$); the circled markers in the acceptance region mark the events which had at least some overlap with the field of view of the SD at any time. Right: Preliminary upper limits on the spectral fluence of UHE photons at Earth for each of the selected gravitational wave sources; the uncertainty bars include both the directional uncertainty of the gravitational wave event (blue) and the uncertainty due to the choice of the spectral index used to calculate the spectral fluence (red); for the second event, the uncertainty bars extend beyond the plotted range, since this source is located right at the edge of the field of view of the Pierre Auger Observatory. For more details, see~\cite{ruehl:2021rmn}.}
\label{fig:gwfollowup}
\end{figure}

Since the first direct detection of gravitational waves in 2015~\cite{LIGOScientific:2016aoc}, the field of multimessenger astronomy has made tremendous progress. In the past years, the transient sources of gravitational wave events---compact binary mergers of black holes and/or neutron stars---have been analyzed by various astronomical instruments. With its unique exposure to UHE particles, the Pierre Auger Observatory has joined the global multimessenger campaign by searching for UHE neutral particles, in particular with follow-up searches for neutrinos (see, e.g.,~\cite{PierreAuger:2016efk}) and photons~\cite{ruehl:2021rmn} in association with gravitational wave events. In this context, the search for photons poses several challenges. Not only is the possible flux of UHE photons from any distant source expected to be heavily attenuated due to interactions with the cosmic background radiation fields, but also the non-negligible background of air-shower events with hadronic origin makes the unambiguous identification of primary photons challenging. The identification of photon candidate events is based on the standard search for photons at the Pierre Auger Observatory using the SD (see Sec.~\ref{sec:sd}). In order to still maintain a high sensitivity towards a possible photon signal from a transient source despite the considerable background, a dedicated gravitational wave selection strategy has been developed which accepts only close or well-localized sources. Three classes of accepted gravitational wave events are defined in Fig.~\ref{fig:gwfollowup}, left, in the space of the $\unit[50]{\%}$ sky localization region and the luminosity distance. While close sources are the most promising candidates to yield a detectable flux of UHE photons, the detection of a photon in coincidence with a distant but well localized source would give a strong hint towards new physics. Out of all gravitational wave events published in the GWTC-1 and -2 catalogs~\cite{LIGOScientific:2018mvr,LIGOScientific:2020ibl}, four events---including GW170817, which originated from the merger of two neutron stars ~\cite{LIGOScientific:2017vwq} and after which a short gamma-ray burst was observed from the same direction in the sky~\cite{LIGOScientific:2017ync}---were selected and analyzed for coincident UHE photons within the time period of one sidereal day after the gravitational wave event. No photon candidate events could be identified. Preliminary upper limits on the spectral fluence within the respective time windows are shown in Fig.~\ref{fig:gwfollowup}, right.\\

A similar analysis ansatz can be used to search for UHE photons in association with other transient events, like flares of the blazar TXS 0506+056~\cite{PierreAuger:2020llu}. Neither during the first period of enhanced neutrino activity observed by IceCube
from October 2014 to February 2015 nor during the second from March 2017 to September
2017~\cite{IceCube:2018dnn} could coincident photon events be identified in the dataset collected with the SD. However, at $\unit{Gpc}$ scales, no flux of UHE photons could possibly be detected at Earth without new physics processes altering the attenuation of photons in the extragalactic medium.


\section{Outlook}
\label{sec:outlook}

Future improvements to the searches for ultra-high-energy photons summarized in this review can naturally be expected from using larger datasets, profiting from the constant increase in exposure over time with which the upper limits scale inversely. Assuming an increase by a factor of two in exposure, this would translate directly into upper limits that are lower by the same factor of two, in the absence of photon candidate events. In addition, the ongoing detector upgrade of the Pierre Auger Observatory, dubbed AugerPrime~\citep{Castellina:2019irv,PierreAuger:2016qzd} will play a major role in the future. A key part of this upgrade is the installation of scintillation detectors on top of the water-Cherenkov detector stations of the SD. The current photon searches discussed before  already exploit the well-known differences in these components between photon- and hadron-induced air showers (see Sec.~\ref{sec:photonshowers}), albeit in a rather indirect way. AugerPrime will allow for a more direct access, which will lead to an overall better separation between photon-induced air showers and the vast hadronic background. In addition, the detector stations will be equipped with radio antennas to measure of the radio signals emitted by an air shower, which act as a proxy to the electromagnetic component and can therefore also be exploited in searches for UHE photons. All of these efforts combined will significantly improve the upper limits on the incoming photon flux or, in the best case, lead to the first unambiguous detection of photons at ultra-high energies.


\section*{Acknowledgements}


\begin{sloppypar}
The successful installation, commissioning, and operation of the Pierre
Auger Observatory would not have been possible without the strong
commitment and effort from the technical and administrative staff in
Malarg\"ue. We are very grateful to the following agencies and
organizations for financial support:
\end{sloppypar}

\begin{sloppypar}
Argentina -- Comisi\'on Nacional de Energ\'\i{}a At\'omica; Agencia Nacional de
Promoci\'on Cient\'\i{}fica y Tecnol\'ogica (ANPCyT); Consejo Nacional de
Investigaciones Cient\'\i{}ficas y T\'ecnicas (CONICET); Gobierno de la
Provincia de Mendoza; Municipalidad de Malarg\"ue; NDM Holdings and Valle
Las Le\~nas; in gratitude for their continuing cooperation over land
access; Australia -- the Australian Research Council; Belgium -- Fonds
de la Recherche Scientifique (FNRS); Research Foundation Flanders (FWO);
Brazil -- Conselho Nacional de Desenvolvimento Cient\'\i{}fico e Tecnol\'ogico
(CNPq); Financiadora de Estudos e Projetos (FINEP); Funda\c{c}\~ao de Amparo \`a
Pesquisa do Estado de Rio de Janeiro (FAPERJ); S\~ao Paulo Research
Foundation (FAPESP) Grants No.~2019/10151-2, No.~2010/07359-6 and
No.~1999/05404-3; Minist\'erio da Ci\^encia, Tecnologia, Inova\c{c}\~oes e
Comunica\c{c}\~oes (MCTIC); Czech Republic -- Grant No.~MSMT CR LTT18004,
LM2015038, LM2018102, CZ.02.1.01/0.0/0.0/16{\textunderscore}013/0001402,
CZ.02.1.01/0.0/0.0/18{\textunderscore}046/0016010 and
CZ.02.1.01/0.0/0.0/17{\textunderscore}049/0008422; France -- Centre de Calcul
IN2P3/CNRS; Centre National de la Recherche Scientifique (CNRS); Conseil
R\'egional Ile-de-France; D\'epartement Physique Nucl\'eaire et Corpusculaire
(PNC-IN2P3/CNRS); D\'epartement Sciences de l'Univers (SDU-INSU/CNRS);
Institut Lagrange de Paris (ILP) Grant No.~LABEX ANR-10-LABX-63 within
the Investissements d'Avenir Programme Grant No.~ANR-11-IDEX-0004-02;
Germany -- Bundesministerium f\"ur Bildung und Forschung (BMBF); Deutsche
Forschungsgemeinschaft (DFG); Finanzministerium Baden-W\"urttemberg;
Helmholtz Alliance for Astroparticle Physics (HAP);
Helmholtz-Gemeinschaft Deutscher Forschungszentren (HGF); Ministerium
f\"ur Kultur und Wissenschaft des Landes Nordrhein-Westfalen; Ministerium
f\"ur Wissenschaft, Forschung und Kunst des Landes Baden-W\"urttemberg;
Italy -- Istituto Nazionale di Fisica Nucleare (INFN); Istituto
Nazionale di Astrofisica (INAF); Ministero dell'Istruzione,
dell'Universit\'a e della Ricerca (MIUR); CETEMPS Center of Excellence;
Ministero degli Affari Esteri (MAE); M\'exico -- Consejo Nacional de
Ciencia y Tecnolog\'\i{}a (CONACYT) No.~167733; Universidad Nacional Aut\'onoma
de M\'exico (UNAM); PAPIIT DGAPA-UNAM; The Netherlands -- Ministry of
Education, Culture and Science; Netherlands Organisation for Scientific
Research (NWO); Dutch national e-infrastructure with the support of SURF
Cooperative; Poland -- Ministry of Education and Science, grant
No.~DIR/WK/2018/11; National Science Centre, Grants
No.~2016/22/M/ST9/00198, 2016/23/B/ST9/01635, and 2020/39/B/ST9/01398;
Portugal -- Portuguese national funds and FEDER funds within Programa
Operacional Factores de Competitividade through Funda\c{c}\~ao para a Ci\^encia
e a Tecnologia (COMPETE); Romania -- Ministry of Research, Innovation
and Digitization, CNCS/CCCDI UEFISCDI, grant no. PN19150201/16N/2019 and
PN1906010 within the National Nucleus Program, and projects number
TE128, PN-III-P1-1.1-TE-2021-0924/TE57/2022 and PED289, within PNCDI
III; Slovenia -- Slovenian Research Agency, grants P1-0031, P1-0385,
I0-0033, N1-0111; Spain -- Ministerio de Econom\'\i{}a, Industria y
Competitividad (FPA2017-85114-P and PID2019-104676GB-C32), Xunta de
Galicia (ED431C 2017/07), Junta de Andaluc\'\i{}a (SOMM17/6104/UGR,
P18-FR-4314) Feder Funds, RENATA Red Nacional Tem\'atica de
Astropart\'\i{}culas (FPA2015-68783-REDT) and Mar\'\i{}a de Maeztu Unit of
Excellence (MDM-2016-0692); USA -- Department of Energy, Contracts
No.~DE-AC02-07CH11359, No.~DE-FR02-04ER41300, No.~DE-FG02-99ER41107 and
No.~DE-SC0011689; National Science Foundation, Grant No.~0450696; The
Grainger Foundation; Marie Curie-IRSES/EPLANET; European Particle
Physics Latin American Network; and UNESCO.
\end{sloppypar}




\bibliographystyle{mdpi}
\bibliography{references}

\end{document}